\documentclass{copernicus}

\begin{document}

\title{Discussion on common errors in analyzing sea level accelerations, solar trends and global warming}

\author{Nicola Scafetta}

\affil{Active Cavity Radiometer Irradiance Monitor (ACRIM) Lab, Coronado, CA 92118, USA}
\affil{Duke University, Durham, NC 27708, USA}

\runningtitle{Common errors in analyzing sea level accelerations, solar trends and global warming}

\runningauthor{N.~Scafetta}

\correspondence{N.~Scafetta (nicola.scafetta@gmail.com)}

\maketitle

%\firstpage{1}

\abstract{ Herein I discuss common errors in applying regression models and
wavelet filters used to analyze geophysical signals. I demonstrate that:
(1)~multidecadal natural oscillations (e.g. the quasi 60\,yr Multidecadal
Atlantic Oscillation (AMO), North Atlantic Oscillation (NAO) and Pacific
Decadal Oscillation (PDO)) need to be taken into account for properly
quantifying anomalous background accelerations in tide gauge records such as
in New York City; (2)~uncertainties and multicollinearity among climate
forcing functions also prevent a proper evaluation of the solar contribution
to the 20th century global surface temperature warming using overloaded
linear regression models during the 1900--2000 period alone; (3)~when
periodic wavelet filters, which require that a record is pre-processed with a reflection methodology, are improperly applied to decompose non-stationary solar and
climatic time series, Gibbs boundary artifacts emerge yielding misleading
physical interpretations. By correcting these errors and using optimized
regression models that reduce multicollinearity artifacts, I found the
following results: (1)~the relative sea level in New York City is not
accelerating in an alarming way, and may increase by about $350\pm30$\,mm
from 2000 to 2100 instead of the previously projected values varying from
$1130\pm480$\,mm to $1550\pm400$\,mm estimated using the methods proposed,
e.g., by \citet{Sallenger} and \citet{Boon}, respectively; (2)~the solar
activity increase during the 20th century contributed at least about 50\,\%
of the 0.8\,$^\circ$C global warming observed during the 20th century instead
of only 7--10\,\% \citep[e.g.:][]{IPCC2007,BS09,Lean2009,Rohde}. The first
result was obtained by using a quadratic polynomial function plus a 60\,yr
harmonic to fit a required 110\,yr-long sea level record. The second result
was obtained by using solar, volcano, greenhouse gases and aerosol
constructors to fit modern paleoclimatic temperature reconstructions
\citep[e.g.:][]{Moberg,Mann2008,Christiansen} since the Medieval Warm Period,
which show a large millennial cycle that is well correlated to the millennial
solar cycle \citep[e.g.: ][]{Kirkby,Scafetta2007,Scafetta2012c}. These
findings stress the importance of natural oscillations and of the sun to
properly interpret climatic changes.
\begin{center}
$\sim$
\end{center}
Cite this article as: \textbf{Scafetta, N.:  Common errors in analyzing sea level accelerations, solar trends and temperature records. \textit{Pattern Recognition in Physics} 1, 37-58, doi: 10.5194/prp-1-37-2013, 2013.}}

\introduction

Geophysical systems are usually studied by analyzing time series. The purpose
of the analysis is to recognize specific physical patterns and to provide
appropriate physical interpretations. Improper applications of complex
mathematical and statistical methodologies are possible, and can yield
erroneous interpretations. Addressing this issue is important because errors
present in the scientific literature may not be promptly recognized and,
therefore, may propagate misleading scientists and policymakers and,
eventually, delay scientific progress.

Herein I briefly discuss a few important examples found in the geophysical
literature where time series tools of analysis were misapplied. These cases
mostly involve multicollinearity artifacts in linear regression models and
Gibbs artifacts in wavelet filters. The following examples are studied:
(1)~the necessity of recognizing and taking into account multidecadal natural
oscillations for properly quantifying anomalous accelerations in tide gauge
records; (2)~the risk of improperly using overloaded multilinear regression
models to interpret global surface temperature records; (3)~how to recognize
Gibbs boundary artifacts that can emerge when periodic wavelet filters are
improperly applied to decompose non-stationary geophysical time series. The
proposed reanalyses correct a number of erroneous interpretations while
stressing the importance of natural oscillations and of the sun to properly
interpret climatic changes.

%f1
\begin{figure*}[t]
\center
\includegraphics[width=13cm]{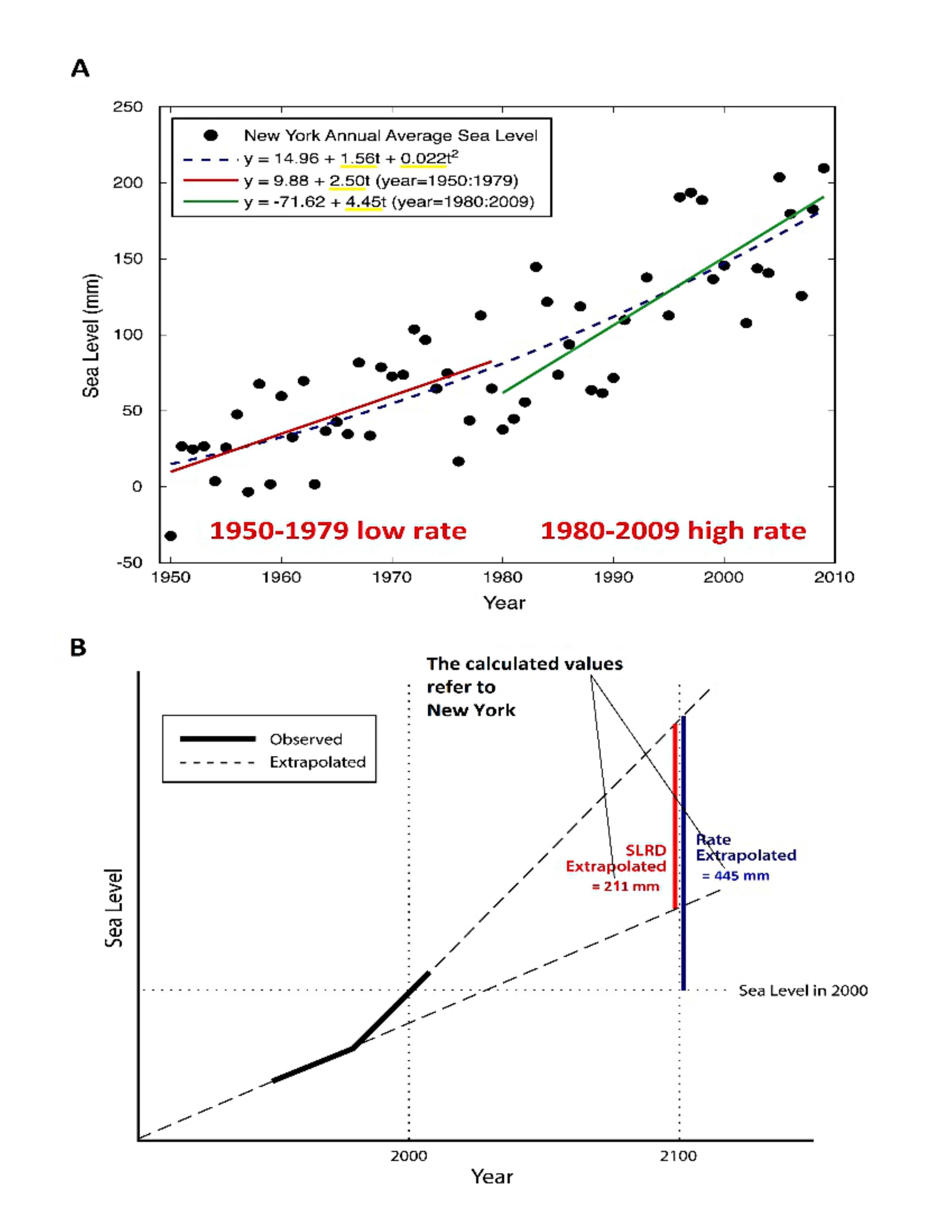}
\caption{Reproduction and comments of \citet{Sallenger}'s figures S7 and S8.
\textbf{(A)}~Sea level record in New York as interpreted in \citet{Sallenger}'s
figures~S7 in their supplementary file. \textbf{(B)}~Predicted sea level rate
difference between the two half-window series (SLRD) for the 21st century as
interpreted  in \citet{Sallenger}'s figures S8 in their supplementary file.}
\end{figure*}

\section{Sea level accelerations versus 60-year oscillations: the New York City case}

Tide gauge records are characterized by complex dynamics driven by different
forces that on multidecadal and multisecular scales are regulated by a
combination of ocean dynamics, of eustasy, isostasy and subsidence
mechanisms, and of global warming
\citep{Boon,Jevrejeva,Morner,Morner2013,Sallenger}. Understanding these
dynamics and correctly quantifying accelerations in tide gauge records is
important for numerous civil purposes. However, changes of rate due to
specific multidecadal natural oscillations should be recognized and separated
from a background acceleration that may be potentially induced by alternative
factors such as anthropogenic global warming. Let us discuss an important
example where this physical aspect was apparently not properly recognized by \citet{Sallenger} and \citet{Boon}.

Figure~1a and~b reproduce (with a few additional comments) figures
S7 and S8 of the supplementary information file published in \citet{Sallenger},
indicated herein as Sa2012. Sa2012's method for interpreting tide
gauge records is detailed below. The example uses the New York City
(NYC) (the Battery) annual average tide gauge record that can be downloaded
from the Permanent Service for Mean Sea Level (PSMSL) (\url{http://www.psmsl.org/})
\citep{Woodworth}.

As Fig.~1a shows, Sa2012 analyzed the tide gauge record for NYC from 1950 to
2009; note, however, that Sa2012's choice appears already surprising because
these data have been available since 1856. Sa2012 linearly fit the periods
1950--1979 and 1980--2009, and found that during 1950--1979 the sea level
rose with a rate of $2.5\pm0.6$\,mm\,yr$^{-1}$, while during 1980--2009 the
rate increased to $4.45\pm0.72$\,mm\,yr$^{-1}$. Thus, a strong apparent
acceleration was discovered and was interpreted as due to the anthropogenic
warming of the last 40\,yr, which could have caused a significant change in
the strength of the Atlantic Meridional overturning circulation and of the
Gulf Stream. This acceleration was more conveniently calculated by fitting
the 1950--2009 period with a second order polynomial, e.g.:
\begin{equation}
g(t)=\frac{1}{2}a(t-2000)^{2}+v(t-2000)+c.\label{eq:1}
\end{equation}
For NYC a 1950--2009 acceleration of $a=0.044\pm0.030$\,mm\,yr$^{-2}$ was
found. Then, Sa2012 repeated the quadratic fit to evaluate the acceleration
during the periods 1960--2009 and 1970--2009, and for NYC the results would
be $a=0.083\pm0.049$\,mm\,yr$^{-2}$ and $a=0.133\pm0.093$\,mm\,yr$^{-2}$,
respectively.  Similarly, \citet{Boon} fit the period from 1969 to 2011 and found $a = 0.20\pm0.07$ mm yr$^{-2}$.

Thus, in NYC not only would the sea level be alarmingly
accelerating, but the acceleration itself has also incrementally increased
during the last decades. Similar results were claimed for other Atlantic
coast cities of North America. Finally, as shown in Fig.~1b, Sa2012
extrapolated its fit curves to 2100 and calculated the sea level rate
difference (SLRD) to provide a first approximation estimate of the
anthropogenic global warming effect on the sea level rise during the 21st
century. For NYC, SLRD would be $\sim211$\,mm if the 1950--1979 and
1980--2009 linear extrapolated trends (reported in the insert of Fig.~1a)
were used, but SLRD would increase to about $\sim890$\,mm if the 1950--1979
linear trend was compared against the 1970--2009 quadratic polynomial fit
extrapolation. Alternatively, by also taking into account the statistical
uncertainty in the regression coefficients, NYC might experience a net sea
level rise of $\sim1130\pm480$\,mm from 2000 to 2100 if Eq.~(\ref{eq:1}) is
used to fit the 1970--2009 period ($a=0.133\pm0.093$\,mm\,yr$^{-2}$,
$v=4.6\pm1.1$\,mm\,yr$^{-1}$, $c=7084\pm8$\,mm) and extrapolated to 2100.

%f2
\begin{figure*}[t]
\center
\includegraphics[width=15cm]{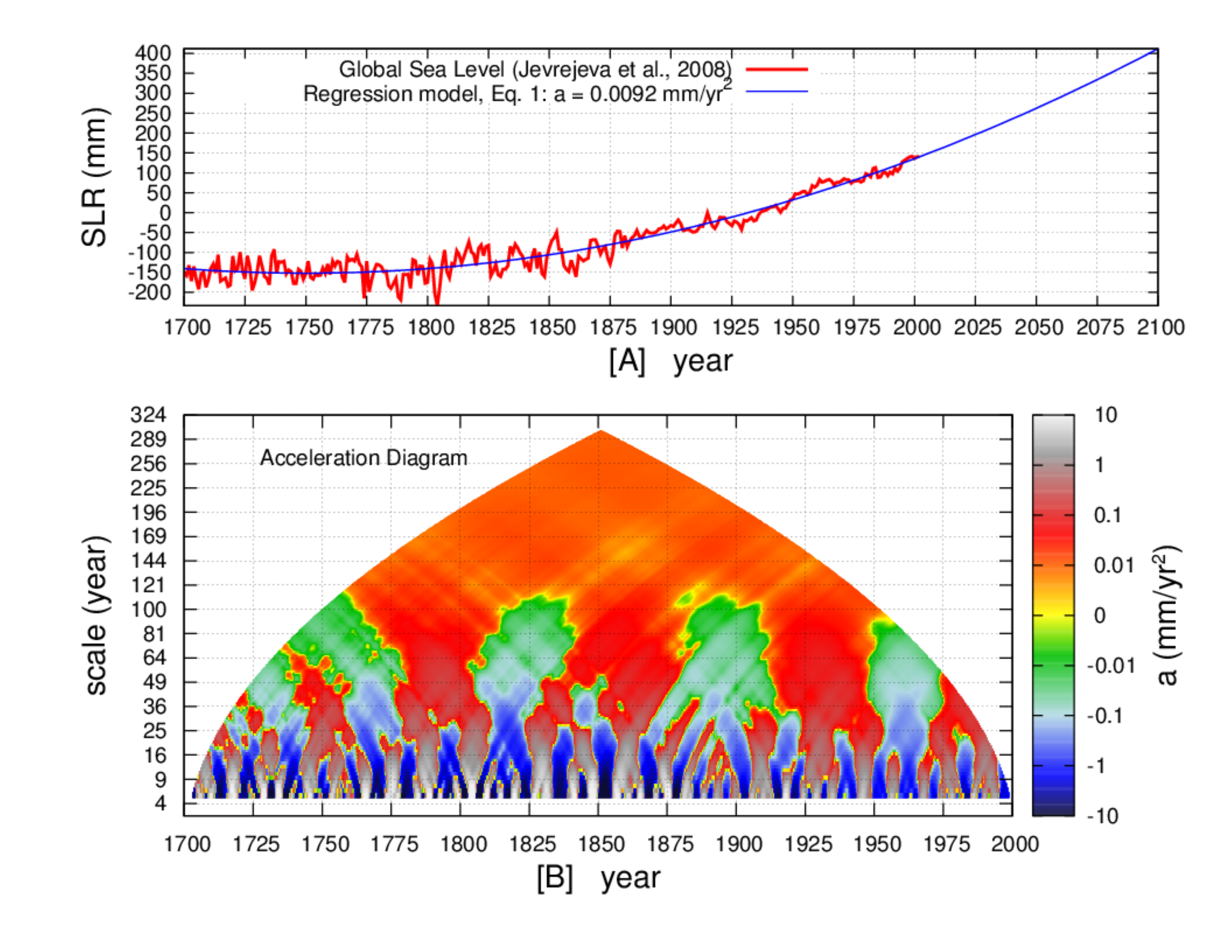}
\caption{\textbf{(A)}~Estimate of the global sea level rise \citep{Jevrejeva} fit
with Eq.~(\ref{eq:1}) (blue). \textbf{(B)}~Scale-by-scale acceleration diagram
of \textbf{(B)}~demonstrating large quasi 60--70\,yr oscillations manifested
by the alternating positive (red area) and negative (green area) accelerations
at scales up to 110\,yr.}
\end{figure*}

%f3
\begin{figure*}[t]
\center
\includegraphics[width=13cm]{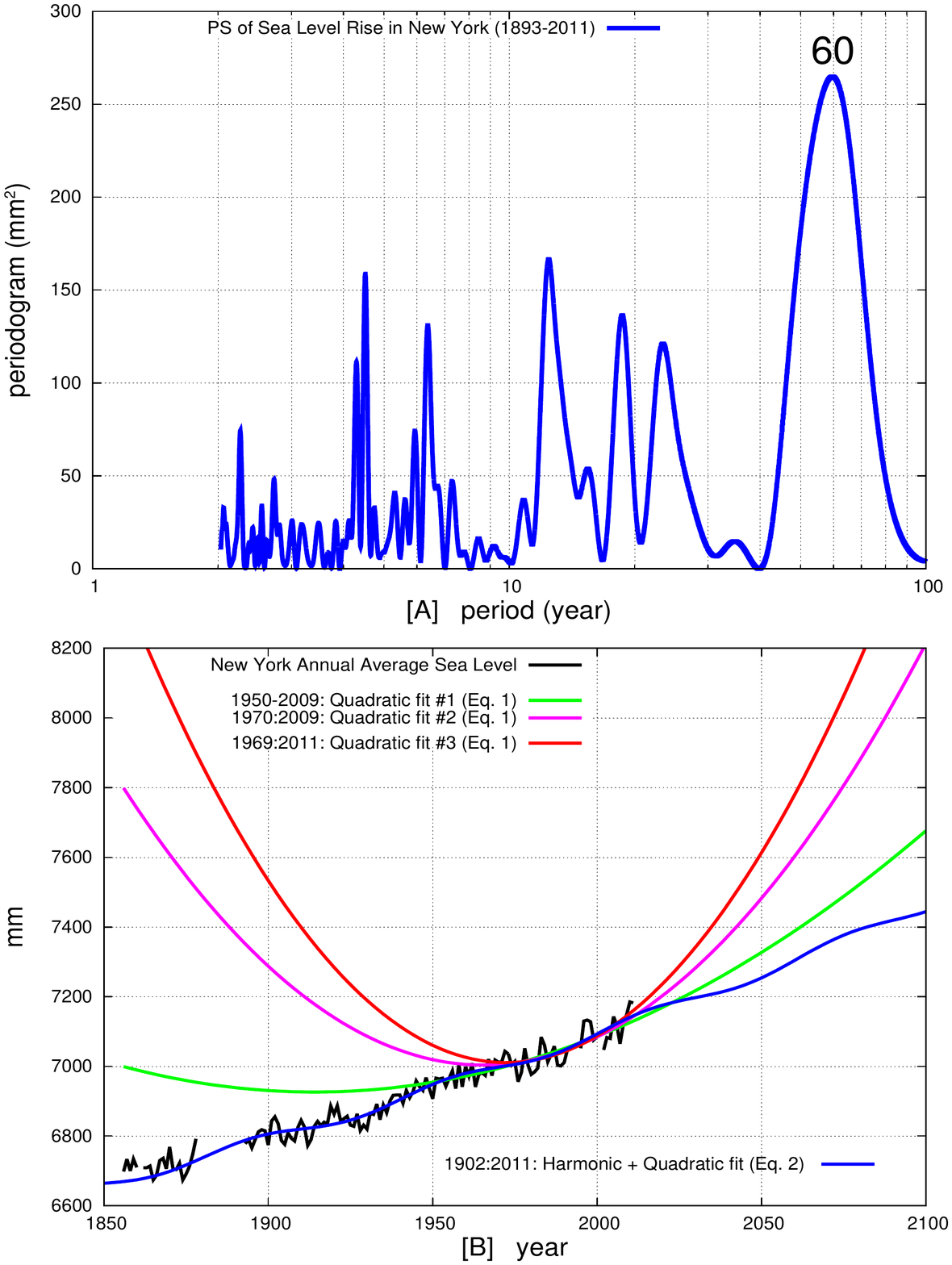}
\caption{\textbf{(A)}~Periodogram of the tide gauge record for New York City
(1893--2011) that highlights a dominant quasi 60\,yr oscillation. The data
are linearly detrended before applying the periodogram algorithm for improved
stability at lower frequencies. \textbf{(B)}~Sea level record for New York
City (black) fitted with Eq.~(\ref{eq:2}) (blue) from 1902 to 2011, and with
the Eq.~(\ref{eq:1}) from 1950 to 2009 (green), from 1970 to 2009 (purple)
and from 1969 to 2011 (red). Projections \#1 and \#2 use \citet{Sallenger}'s method,
projection \#3 uses Boon's~(\citeyear{Boon}) method. The blue model agrees far better
with the data since 1856 and likely produces the most realistic projection
for the 21st century; see also \citet{Scafetta2013b} for additional details.
}
\end{figure*}

However, Sa2012's result does not appear robust because, as I will
demonstrate below, the geometrical convexity observed in the NYC tide gauge
record from 1950 to 2009 was very likely mostly induced by a quasi 60\,yr
oscillation that is already known to exist in the climate system. In fact,
numerous ocean indexes such as the Multidecadal Atlantic Oscillation (AMO),
the North Atlantic Oscillation (NAO) and the Pacific Decadal Oscillation
(PDO) oscillate with a quasi 60\,yr period for centuries and millennia
\citep[e.g.:][]{Morner1989,Morner(1990),Klyashtorin,Mazzarella,Knudsen,Scafetta2013b},
as well as global surface temperature records
\citep[e.g.:][]{Kobashi,Qian,Scafetta2010,Scafetta2012a,Schulz}. In
particular, \citet{Scafetta2010,Scafetta2012c,Scafetta2012d} provided
empirical and theoretical evidence that the observed multidecadal oscillation
could be solar/astronomical-induced, could be about 60\,yr-long from 1850 to
2012 and could be modulated by other quasi-secular oscillations \citep[e.g.:
][]{Ogurtsov,Scafetta2012c,Scafetta2013}. In fact, a quasi 60\,yr oscillation
is particularly evident in the global temperature records since 1850:
1850--1880, 1910--1940 and 1970--2000 were warming periods and 1880--1910,
1940--1970 and 2000--(2030?) were cooling periods. This quasi 60\,yr
oscillation is superposed to a background warming trend which may be due to
multiple causes (e.g.: solar activity, anthropogenic forcings and urban heat
island effects) \citep[e.g.:
][]{Scafetta2007,Scafetta2009b,Scafetta2010,Scafetta2012a,Scafetta2012b,Scafetta2012c}.
Because the climate system is evidently characterized by numerous
oscillations, tide gauge records could be characterized by equivalent
oscillations too.

%f4
\begin{figure*}
\center
\includegraphics[width=15cm]{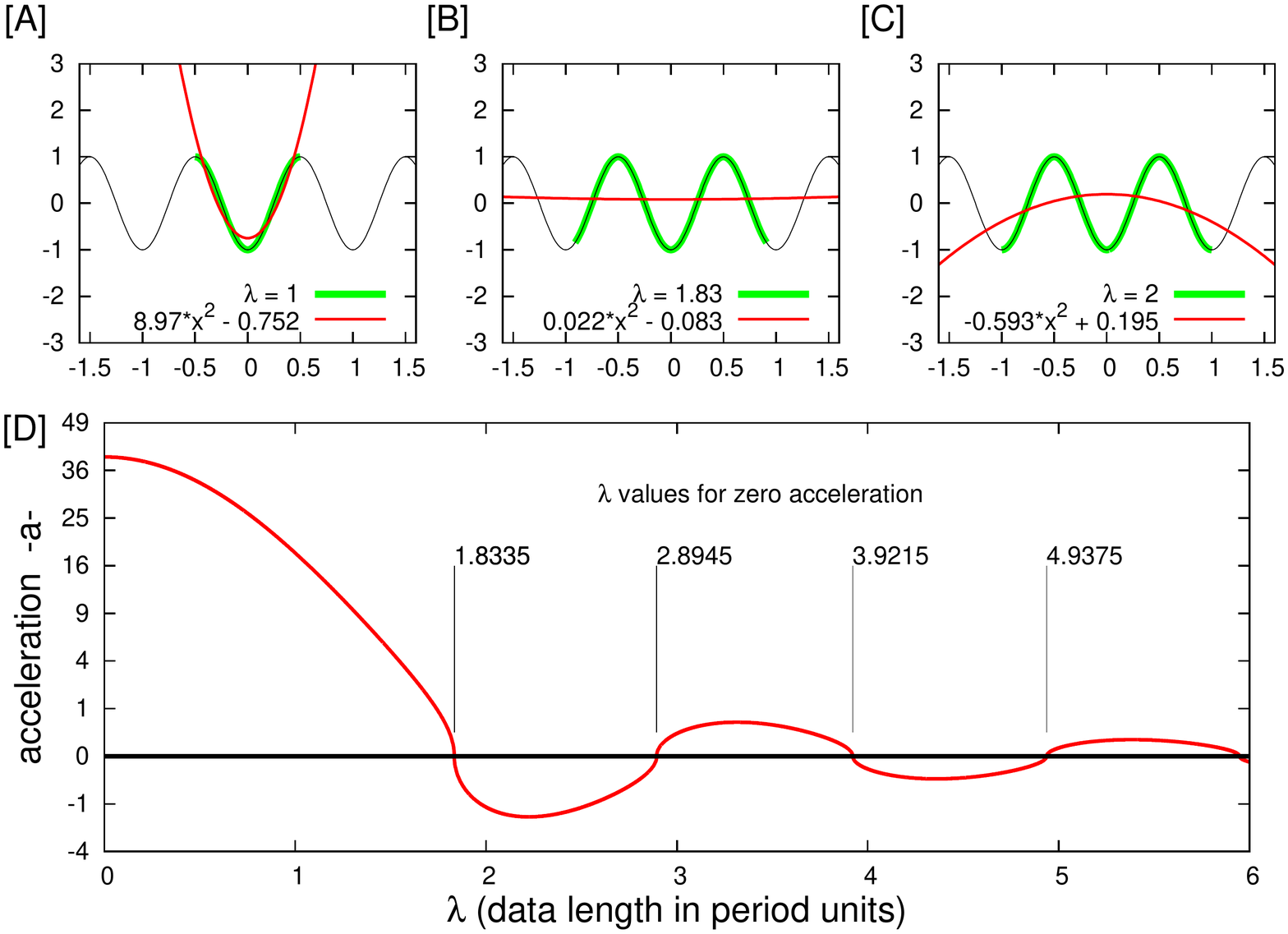}
\caption{\textbf{(A)}, \textbf{(B)} and \textbf{(C)}~show a stationary
harmonic signal (black) of unit period fit with Eq.~(\ref{eq:1}) (red) using
three different records length (green), $\lambda= 1$, $1.83$ and $2$,
respectively. \textbf{(D)}~The regression acceleration coefficient in
function of the record length $\lambda$. The figure highlights the values of
$\lambda$ that make the acceleration $a=0$, indicating regression
orthogonality between the harmonic signal and the quadratic polynomial. From
\citet{Scafetta2013b}.}
\end{figure*}

Indeed, a quasi 60\,yr oscillation has been found in numerous sea level
records since 1700 \citep{Chambers,Jevrejeva,Parker}. Figure~2a shows the
global sea level record from 1700 to 2000 proposed by \citet{Jevrejeva} fit
with Eq.~(\ref{eq:1}) ($a=0.0092\pm0.0004$\,mm\,yr$^{-2}$;
$v=2.31\pm0.06$\,mm\,yr$^{-1}$; $c=136\pm4$\,mm). In addition to a relatively
small acceleration since 1700\,AD, which, if continues, will cause a global
sea level rise of about $277\pm8$\,mm from 2000 to 2100, the global sea level
record clearly presents large 60--70\,yr oscillations. This is better
demonstrated in Fig.~2b that shows the scale-by-scale palette acceleration
diagram of this global sea level record \citep{Jevrejeva,Scafetta2013b}. Here
the color of a dot at coordinate (x, y) indicates the acceleration $a$
(calculated with Eq.~ \ref{eq:1}) of a $y$-year-long interval centered in $x$.
The color of the dot at the top of the diagram, which in this case is
approximately orange, indicates the global acceleration for the 1700--2000
period, $a=0.0092\pm0.0004$\,mm\,yr$^{-2}$. The diagram also suggests that
for scales larger than 110\,yr the acceleration is almost homogeneous, around
$0.01$\,mm\,yr$^{-2}$ or less (orange/yellow color) at all scales and times
\citep{Scafetta2013b}. For example, during the preindustrial 1700--1900
period $a=0.009\pm0.001$\,mm\,yr$^{-2}$; during the industrial 1900--2000
period $a=0.010\pm0.0004$\,mm\,yr$^{-2}$. Thus, the observed acceleration
appears to be independent of the 20th century anthropogenic global warming
and could be a consequence of other phenomena, such as the quasi-millennial
solar/climate cycle \citep{Bond,Kerr,Kobashi2013,Kirkby,Scafetta2012c}
observed throughout the Holocene. The millennial solar/climatic cycle has
been in its warming phase since 1700, which characterized the Maunder solar
minimum during the Little Ice Age. Finally, the alternating quasi regular
large green and red areas evident at scales from 30 to 110\,yr indicate a
change of acceleration (from negative to positive, and vice-versa) that
reveals the existence of a quasi 60--70\,yr oscillation since 1700. Strong
quasi decadal and bidecadal oscillations are observed at scales below 30\,yr.
In conclusion, because the global sea level record presents a clear quasi
60\,yr oscillation that also well correlates with the quasi 60\,yr
oscillation found in the NAO index since 1700 \citep{Scafetta2013b}, there is
the need to check whether the tide gauge record of NYC too may have been
affected by a quasi 60\,yr oscillation.

Herein I extend the finding discussed in \citet{Scafetta2013b}. Figure~3a
shows the periodogram of the tide gauge record for NYC from 1893 to 2011: the
data available before 1893 are excluded from the analysis because the record
is seriously incomplete. The periodogram is calculated after the three
missing years (in 1992, 1994 and 2001) are linearly interpolated, and the
linear trend ($y(t)=2.98(t-2000)+7088$) is detrended because the periodogram
gives optimal results if the time series is stationary. The spectral analysis
clearly highlights, among other minor spectral peaks, a dominant frequency at
a period of about 60\,yr, which is a typical major multidecadal oscillation
found in PDO, AMO and NAO indexes
\citep{Klyashtorin,Knudsen,Mazzarella,Scafetta2012a,Scafetta2013b,Manzi}.
This quasi 60\,yr oscillation, after all, is clearly visible in the NYC tide
gauge record once this record is plotted since 1856, as shown in Fig.~3b.

Consequently, for detecting a possible background sea level acceleration
for NYC there is a need of adopting an upgraded regression model that
at least must be made of a harmonic component plus a quadratic function
of the type:
\begin{equation}
f(t)=H\cos\left(2\pi\frac{t-T}{60}\right)+\frac{1}{2}a(t-2000)^{2}+v(t-2000)+c.\label{eq:2}
\end{equation}
Other longer multisecular and millennial oscillations may be added to the
model \citep{Bond,Kerr,Ogurtsov,Qian,Scafetta2012c,Schulz} but, because only
about one century of data are herein analyzed, Eq.~(\ref{eq:2}) cannot be
expanded.

To determine the exact length of the time period required to avoid
multicollinearity and make the 60\,yr oscillation orthogonal to the quadratic
polynomial term, a test proposed in \citet{Scafetta2013b} is herein
rediscussed for the benefit of the reader. Figure~4a, b and c show fits of a
periodic signal of unit period 1 of different length with a quadratic
polynomial: the acceleration clearly varies in function of the length of the
record $\lambda$. Figure~4d shows that the acceleration $a$ oscillates around
zero in function of $\lambda$. The minimum length that makes the acceleration
zero is $\lambda=1.8335$ times the length of the period of the oscillation.
Thus, to optimally separate a 60\,yr oscillation from a background
acceleration, there is the need of using a $1.8335\times 60=110$\,yr-long
sequence. Indeed, as Fig.~2b shows, the alternation between the red and the
green areas ends at scales close to $110$\,yr indicating that there is the
need of using more than 100\,yr for filtering a background acceleration out
from the quasi 60\,yr oscillation.

Note that Sa2012's regression model was applied to 60\,yr-long and shorter
intervals from 1950 to 2009. As Fig.~4a and~d clearly show, using 60\,yr-long
and shorter records ($\lambda\leq1$ period of the oscillation), makes the
regression model unable to separate a background acceleration from a 60\,yr
oscillation because the two curves are significantly collinear, and a strong
acceleration simply related to the bending of the 60\,yr oscillation would be
found. In the next section, the multicollinearity problem in regression
models will be discussed more extensively.

NYC sea level data have been intermittently available since 1856, but as of
1893 only three annual means are missing, so the model given by
Eq.~(\ref{eq:2}) can be tested for this record because it is about 120\,yr
long from 1893 to 2011. Figure~3b shows
the sea level record for NYC since 1856 (black) fitted with Eq.~(\ref{eq:2})
for the required optimal 110\,yr interval from 1902 to 2011 (blue). The fit
gives $H=16\pm4$\,mm; $T=1956\pm2.5$\,yr; $a=0.006\pm0.005$\,mm\,yr$^{-2}$;
$v=3.3\pm0.3$\,mm\,yr$^{-1}$; $c=7094\pm6$\,mm. For comparison, Fig.~3b also
shows Sa2012's and Boon's~(\citeyear{Boon}) methods using Eq.~(\ref{eq:1}): (1)~the fit
is done from 1950 to 2009 (green) ($a=0.044\pm0.030$\,mm\,yr$^{-2}$,
$v=3.7\pm0.7$\,mm\,yr$^{-1}$, $c=7086\pm7$\,mm); (2)~the fit is done from
1970 to 2009 (purple) ($a=0.133\pm0.1$\,mm\,yr$^{-2}$,
$v=4.6\pm1.1$\,mm\,yr$^{-1}$, $c=7084\pm8$\,mm); (3)~the fit is done from
1969 to 2011 (red) ($a=0.20\pm0.07$\,mm\,yr$^{-2}$,
$v=5.5\pm0.9$\,mm\,yr$^{-1}$, $c=7087\pm7$\,mm). Projections \#1 and \#2 use
Sa2012's method, projection \#3 uses Boon's~(\citeyear{Boon}) method.

To test the sufficient stability of my result, the analysis for the two
non-overlapping periods 1856--1934 and 1934--2012 is repeated. In the first
case the fit gives $H=14\pm6$\,mm; $T=1963\pm5$\,yr;
$a=0.018\pm0.023$\,mm\,yr$^{-2}$; $v=4.1\pm2.4$\,mm\,yr$^{-1}$;
$c=7107\pm122$\,mm. In the second case the fit gives $H=16\pm5$\,mm;
$T=1957\pm5$\,yr; $a=0.015\pm0.027$\,mm\,yr$^{-2}$;
$v=3.6\pm0.8$\,mm\,yr$^{-1}$; $c=7095\pm7$\,mm. Because in the three cases
the correspondent regression values are compatible to each other within their
uncertainty and the regression model calibrated from 1856 to 1934 hindcasts
the data from 1934 to 2012, and vice versa, the regression model,
Eq.~(\ref{eq:2}), can be considered sufficiently stable for interpreting the
available data.

On the contrary, using Eq.~(\ref{eq:1}) to fit 60\,yr periods it is
obtained: (1)~from 1890 to 1949, $a=0.091\pm0.027$\,mm\,yr$^{-2}$;
(2)~from 1920 to 1979, $a=-0.043\pm0.025$\,mm\,yr$^{-2}$; (3)~from
1950 to 2009, $a=0.044\pm0.030$\,mm\,yr$^{-2}$. Because the acceleration
values of the three 60\,yr sub-periods are not compatible to each
other within their uncertainty, the regression model Eq.~(\ref{eq:1})
does not capture the dynamics of the available NYC sea level data.
However, the absolute values of the three accelerations are compatible
to each other. Thus, the 1950--2009 acceleration value, $a=0.044\pm0.03$\,mm\,yr$^{-2}$,
does not appear to be anomalous, but it is well within
the natural variability of a system that oscillates with a quasi 60\,yr
cycle around a quasi linear upward trend. See \citet{Scafetta2013b}
for additional discussion demonstrating that the accelerations found
using the intervals proposed by Sa2012 and \citet{Boon} are arbitrary.

Figure~3b also highlights that Eq.~(\ref{eq:2}) hindcasts quite well the
relative sea level in NYC from 1856 to 1901, whose period was not used to
calibrate the regression model, which adopted only data from 1902 to 2011.
Therefore, the model proposed in Eq.~(\ref{eq:2}) reconstructs the available
data since 1856, takes into account an influence of known climatic oscillations (e.g.
the quasi 60\,yr AMO oscillation) and may be reasonably used as a first
approximation forecast tool. On the contrary, Sa2012 and Boon's~(\citeyear{Boon})
models immediately miss the data before 1950, 1969 and 1970, respectively,
and ignore the existence of known multidecadal natural oscillations of the
climate system. Consequently, the usefulness of the latter models for
hindcast/forecast purposes should be questioned even on short periods.
Essentially, Sa2012 Boon's~(\citeyear{Boon}) methodologies are too simplistic because,
as evident in Fig.~3b, they do not capture the dynamics of the available data
and, consequently, miss the true dynamical properties of the system.

As Fig.~3b shows, the adoption of Eq.~(\ref{eq:2}) implies that the relative
sea level in NYC accelerated 7 to 22 times \textit{less} than what was
obtained with Sa2012's quadratic fit alone during the two 30\,yr periods
1950--2009 and 1970--2009, respectively. By using the same extrapolation
methodology proposed in Sa2012 and assuming that Eq.~(\ref{eq:2}) persists
during the 21st century, the relative sea level in NYC could rise about
$350\pm30$\,mm from 2000 to 2100, which is significantly less than what
Sa2012's quadratic model extrapolation would suggest, that is up to about
$1130\pm480$\,mm, or, using Boon's~(\citeyear{Boon}) model, the projected sea level
would be $1550\pm400$\,mm from 2000 to 2100, as Fig.~3b shows.

In conclusion, the convexity of the NYC tide gauge record from 1950 to 2009
was very likely mostly induced by the quasi 60\,yr AMO-NAO oscillation that
strongly influences the Atlantic cost of North America and can be observed
also in the global sea level record since 1700, shown in Fig.~2. However,
Sa2012 mistook the 1950--2009 geometrical convexity of the NYC record as if it were due to an
anomalous acceleration. Evidently, Sa2012's 21st century projections for sea
level rise in numerous locations need to be revised downward by taking it
into account that the known multidecadal variability of the climate system
would imply a significantly lower background acceleration than what they have
estimated. A similar critique applies to the results by \citet{Boon} too, who
also used Eq.~(\ref{eq:1}) to analyze a number of US and Canadian tide gauge
records over a 43\,yr period from 1969 to 2011 and, for NYC, he found a
1969--2011 acceleration of $a=0.20\pm0.07$\,mm\,yr$^{-2}$ and projected an
alarming sea level rise of $570\pm180$\,mm above the 1983--2001 sea level
mean by 2050. On the contrary, other authors
\citep{Houston,Parker,Scafetta2013b} analyzed numerous secular-long tide
gauge records and found small (positive or negative) accelerations close to
zero ($\sim\pm0.01$\,mm\,yr$^{-2}$). Figure~2a shows that a global estimate
of the sea level rise since 1700 presents an acceleration slightly smaller
than $0.01$\,mm\,yr$^{-2}$ since 1700, which may have also been partially
driven by the great millennial solar/climate cycle
\citep{Bond,Kerr,Kirkby,Scafetta2012c}, which will be more extensively
discussed in the next section.

%f5
\begin{figure*}[t]
\center
\includegraphics[width=15cm]{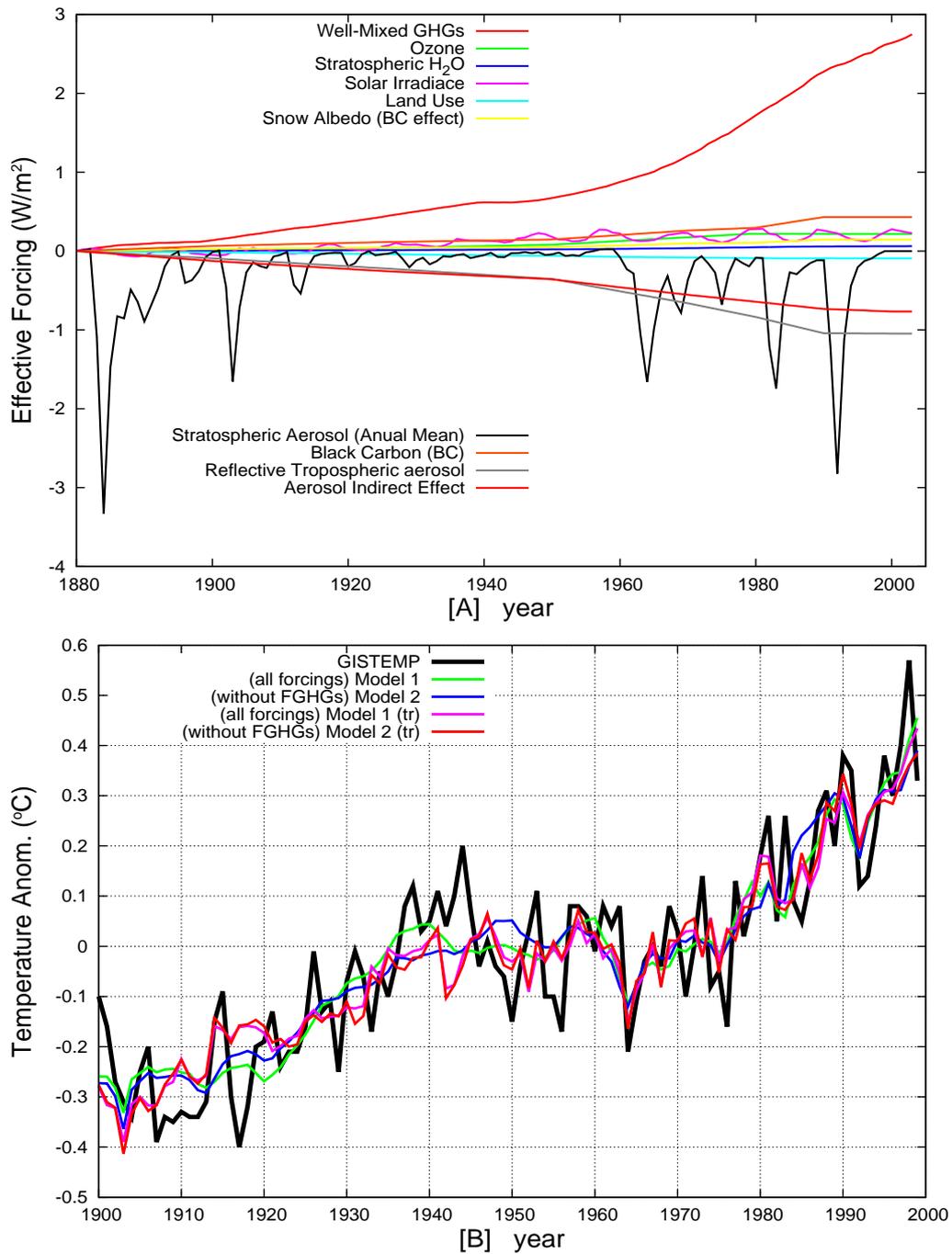}
\caption{\textbf{(A)}~The ten regression constructors used in Eq.~(\ref{eq:3}) by
\citet{BS09}. \textbf{(B)}~Four regression models of the global surface temperature
record (GISTEMP) using all ten constructors (Model 1), and using nine
constructors excluding the well-mixed greenhouse gases (CO$_{2}$ and
CH$_{4}$) (Model 2). The two curves with -tr- adopt forcing functions
truncated to the second decimal digit. See Table 2 for the regression
coefficients.}
\end{figure*}

\section{Multi-linear regression models and the  multicollinearity problem:
estimates of  the solar signature on climatic records}

A number of authors have studied global surface temperature records using
multilinear regression models to identify the relative contribution of known
forcings of the earth's temperature field. For example, \citet{Douglass}, and
\citet{Gleisner} interpreted temperature records for the period 1980--2002
using four regression analysis constructors: an 11\,yr solar cycle signal
without any trend, the volcano signal, the ENSO signal, which captures fast
climatic fluctuations, and a linear trend that can capture everything else
responsible for the 1980--2002 warming trend, including the warming component
induced by anthropogenic greenhouse gases (GHG). The four chosen constructors
are sufficiently geometrically orthogonal and physically independent of each
other. Geometrical orthogonality and physical independence are necessary
conditions for efficiently decomposing a signal using multilinear regression
models. On the contrary, multilinear regression models may produce seriously
misleading and inconclusive results if used with constructors multicollinear
to each other. In fact, it is well known that in presence of
multicollinearity among the regression predictors the estimated regression
coefficients may change quite erratically in response to even minor changes
in the model or the data yielding misleading interpretations.

An improper application of the multilinear regression method is found in
\citet{BS09}, indicated herein as BS09. These authors aimed to demonstrate
that the increased solar activity during the 20th century contributed only
7\,\% of the observed global warming from 1900 to 2000 (about
$0.056$\,$^\circ$C out of a total warming of $0.8$\,$^\circ$C) as commonly
found with general circulation models \citep{Hansen2001,Hansen2007,IPCC2007}.
To do this, BS09 adopted a linear regression model of the global surface
temperature that uses as constructors the 10 forcing functions of the GISS
ModelE \citep{Hansen2007}: these 10 forcing functions are depicted in
Fig.~5a. However, as I will demonstrate below, BS09's approach is neither
appropriate nor sufficiently robust because their chosen constructors do not
satisfy the geometrical orthogonality nor the necessary physical
requirements. This two-fold failure is seen in a number of ways.

The first way BS09 multi-linear regression fails is mathematical.
The predictors of a multilinear regression model must be sufficiently
linearly independent, i.e. it should not be possible to express any
predictor as a linear combination of the others. On the contrary,
all 10 forcing functions used as predictors in BS09, with the exception
of the volcano one, present a quasi monotonic trend (positive or negative)
during the 20th century \citep{Hansen2007}. These smooth trends are
geometrically quite collinear to one other. Thus, these forcing functions
are strongly non-orthogonal and strongly cross-correlated.

This is demonstrated in Table~1 where the cross-correlation coefficients
among the ten forcing functions depicted in Fig.~5a from 1900 to
1999 are reported. The table clearly indicates that with the exception
of the volcano forcing, all other forcing functions are strongly (positively
or negatively) cross-correlated ($|r|>0.65$ for just 100\,yr and
in most cases $|r|>0.95$, which indicates an almost 100\,\% cross-correlation).
The strong cross-correlation among 9 out of 10 constructors makes
BS09 multilinear regression model extremely sensitive to data errors
and to the number of the constructors.

Paradoxically, even a multilinear regression model that does not use
the well-mixed GHG forcing ($F_\mathrm{GHGs}$) at all, which includes also CO$_{2}$
and CH$_{4}$ greenhouse records, would well fit the temperature data
with appropriate regression coefficients in virtue of the extremely
good multicollinearity that the $F_\mathrm{GHGs}$ record has with other eight
forcing functions, as Table~1 shows. This is demonstrated in Fig.~5b
where the GISTEMP global surface temperature record \citep{Hansen2001}
is fit with two multilinear regression models of the type:
\begin{equation}
T(t)=\sum_{i=1}^{N}\beta_{i}F_{i}(t)+c,\label{eq:3}
\end{equation}
where $T(t)$ is the temperature record to be constructed, $\beta_{i}$ are the
linear regression coefficients and $F_{i}(t)$ are the 10 forcing functions
used in BS09. Model 1 uses all ten forcing functions, as used in BS09; Model
2 uses nine forcing functions, where the well-mixed GHG forcing
($F_\mathrm{GHGs}$) is excluded. The regression coefficients of the two
models are reported in Table 2. Moreover, to demonstrate the sensitivity of
the regression algorithm to even small changes of the data, I repeated the
calculation and reported in the last two columns of Table 2 (labeled with
``tr'') the regression coefficients obtained with the same two models, using
forcing functions truncated at 2 decimal digits (the original functions have
4 decimal digits). Figure~5b clearly shows that Model 1 and Model 2, in both
the truncated and untruncated cases, perform almost identically, despite the
fact that individual regression coefficients reported in Table 2 are very
different from each other in the four cases; the statistical errors
associated to these regression coefficients are therefore very large.

%t1
\begin{table*}
\caption{Cross-correlation coefficients among the GISS ModelE forcing functions.
$F_\mathrm{Sun}$ is the solar forcing at the top of the atmosphere, $F_\mathrm{GHG}$
describes radiative forcing due to well-mixed GHG concentrations, $F_\mathrm{O_{3}}$
describes forcing due to stratospheric ozone, $F_\mathrm{H_{2}O}$
represents stratospheric moisture, $F_\mathrm{Refl}$ reflective tropospheric aerosols,
$F_\mathrm{land}$ landscape changes (land use), $F_\mathrm{snow}$ snow albedo
(black carbon effect), $F_\mathrm{Aer}$ represents stratospheric aerosols (volcanoes),
$F_\mathrm{BC}$ is black carbon, $F_\mathrm{AIE}$ is aerosol indirect effect.
Value of the cross-correlation coefficient larger than 0.6 indicates that the
two records are almost 100\,\% correlated. Only the volcano signal is poorly
correlated to the other records.}
\center
\begin{tabular}{lrrrrrrrrrr}
\tophline
        &   $F_\mathrm{GHGs}$ &     $F_\mathrm{O_{_3}}$   &     $F_\mathrm{H_{_2}O}$  &     $F_\mathrm{Sun}$  &     $F_\mathrm{land}$ &     $F_\mathrm{snow}$
        &     $F_\mathrm{Aer}$   &    BC    &      $F_\mathrm{Refl}$      &      AIE  \\
        \hhline
 $F_\mathrm{GHGs}$      &     1     &     0.94  &     0.99  &     0.65  &     $-$0.9  &     0.99  &     $-$0.3  &      0.99 &      $-$0.99      &      $-$0.97      \\
 $F_\mathrm{O_{_3}}$  &     0.94  &     1     &     0.98  &     0.74  &     $-$0.97 &     0.96  &     $-$0.32 &      0.96 &     $-$0.98  &     $-$0.99       \\
 $F_\mathrm{H_{_2}O}$ &     0.99  &     0.98  &     1     &     0.71  &     $-$0.95 &     0.99  &     $-$0.31 &      0.99 &     $-$1     &     $-$1   \\
 $F_\mathrm{Sun}$ &     0.65  &     0.74  &     0.71  &     1     &     $-$0.83 &     0.66  &     $-$0.11 &      0.66 &      $-$0.7  &     $-$0.75      \\
 $F_\mathrm{land}$      &     $-$0.9  &     $-$0.97 &     $-$0.95 &     $-$0.83 &     1     &     $-$0.92 &     0.26  &      $-$0.91      &      0.94  &     0.97 \\
 $F_\mathrm{snow}$      &     0.99  &     0.96  &     0.99  &     0.66  &     $-$0.92 &     1     &     $-$0.34 &      1    &     $-$0.99  &     $-$0.98       \\
 $F_\mathrm{Aer}$ (volcano)   &     $-$0.3  &     $-$0.32 &     $-$0.31 &     $-$0.11 &     0.26  &     $-$0.34 &      1    &     $-$0.33  &     0.32  &    0.3   \\
 BC   &     0.99  &     0.96  &     0.99  &     0.66  &     $-$0.91 &     1     &     $-$0.33 &     1      &    $-$0.99 &      $-$0.98      \\
 $F_\mathrm{Refl}$      &     $-$0.99 &     $-$0.98 &     $-$1    &     $-$0.7  &     0.94  &     $-$0.99 &     0.32  &      $-$0.99      &     1      &     0.99  \\
 AIE  &     $-$0.97 &     $-$0.99 &     $-$1    &     $-$0.75 &     0.97  &     $-$0.98 &     0.3   &     $-$0.98  &    0.99  &      1     \\
 \bottomhline
 \end{tabular}
 \end{table*}

Because it is also possible to equally well reconstruct the temperature
record with Model 2,  the methodology adopted by BS09 could also
be used to demonstrate that the anthropogenic greenhouse gases such as
CO$_{2}$ and CH$_{4}$ are irrelevant for explaining the global warming
observed from 1900 to 2000. Moreover, for physical considerations the
regression coefficients must be positive, but the regression algorithm finds
also negative values, which is another effect of the multicollinearity of the
predictors. This result clearly demonstrates the non-robustness and the
physical irrelevance of the multilinear regression model methodology
implemented in BS09 and, indirectly, also questions their conclusion that the
solar activity increase during the 20th century contributed only $\sim7$\,\%
of the total warming.

The results of the linear regression model used by BS09 would also strongly
depend on the specific total solar irradiance record used as constructor.
BS09 used a model by \citet{Lean2000} which poorly correlates with the
temperature, and they reached a result equivalent to \citet{Lean2009} who
just used an update solar model. However, total solar irradiance records are
highly uncertain and other solar reconstructions \citep[e.g.:][]{Hoyt}
correlate quite better with the temperature records from 1900 to 2000
\citep[e.g.: ][]{Soon2005,Soon2011,Soon2013} and could reconstruct a larger
percentage of the 20th century global warming by better capturing the quasi
60\,yr oscillation found in the temperature records.

However, because the linear regression analysis requires accurate
constructors and the multidecadal patterns of the solar records, as
well as those of the other forcing functions, are highly uncertain,
it is better to use an alternative methodology to test how well the
GISS ModelE simulates the climatic solar signatures. For example,
it is possible to extract the quasi 11\,yr solar cycle signatures
from a set of climatic records and compare the results against the
GISS ModelE predictions. I adopted the method proposed by \citet{Douglass}
and \citet{Gleisner} that uses only four constructors (as already
explained above) for the period 1980--2003:
\begin{equation}
T(t)=\alpha_{V}V(t)+\alpha_{S}S(t)+\alpha_{E}E(t)+a(t-1980)+b.\label{eq:4}
\end{equation}
The function $V(t)$ is the monthly-mean optical thickness at 550\,nm
associated with the volcano signal; $S(t)$ is the 10.7\,cm solar
flux values, which is a good proxy for the 11\,yr modulation of the
solar activity (not for the multi-decadal trend); $E(t)$ is the ENSO
signal (it has been lag-shifted by four months for autocorrelation
reasons also indicated in \citealp{Gleisner}); and the linear trend
captures any linear warming trend the data may present, which may
be due to multiple physical causes such as anthropogenic GHG forcings.

%t2
  \begin{table}
\caption{Multilinear regression coefficients ($^o$C\,m$^2$\,W$^{-1}$) relative to
the ten forcing functions used in \citet{BS09} to fit the GISTEMP dataset from 1900
to 1999. Model 1 uses all ten forcing functions, Model 2 sets the GHGs
forcing equal to \emph{zero} and reconstruct the temperature record with the
other nine forcing functions. The last two columns with -tr- are obtained by
truncating the forcing functions at 2 decimal digits. See Fig.~5b.}
\center
  \begin{tabular}{lrrrr}
  \tophline
        &       Model 1 &     Model 2 &    Model 1      &     Model 2       \\
        &&&(tr)&(tr)\\
        \hhline
$F_\mathrm{GHGs}$ &1.27&0.00& 0.34  &     0.00  \\
$F_\mathrm{O_{_3}}$      &11.0&2.47& $-$5.39 &     $-$7.74 \\
$F_\mathrm{H_{_2}O}$  &10.2&65.2& 2.86  &     3.54  \\
$F_\mathrm{Sun}$     &0.30&0.31& 0.54  &     0.52  \\
$F_\mathrm{land}$ &$-$32.3&$-$13.6&     $-$4.88 &     $-$3.04 \\
$F_\mathrm{snow}$    &$-$10.1&$-$7.57&     7.19  &     7.80  \\
$F_\mathrm{Aer}$ (volcano)     &0.05&0.06& 0.04  &     0.05  \\
$F_\mathrm{BC}$      &13.3&8.58& $-$4.59 &     $-$5.45 \\
$F_\mathrm{Refl}$ &6.12&4.20& $-$0.97 &     $-$2.46 \\
$F_\mathrm{AIE}$     &8.91&4.78& $-$0.38 &     $-$0.68 \\
const ``$c$''   &0.00&$-$0.05&      $-$0.42       &     $-$0.45       \\
\bottomhline
\end{tabular}
\end{table}

%t3
  \begin{table}
  \caption{Cross-correlation coefficients  of the four temperature constructors
  used  for the results depicted in Fig.~6. The four constructors are reasonably orthogonal.}
  \center
  \begin{tabular}{lrrrr}
  \tophline
  &   Volcano     &     Sun   &     ENSO  &     Linear      \\
  \hhline
  Volcano   &     1     &     0.01  &     0.42  &     $-$0.24 \\
  Sun &     0.01  &     1     &     $-$0.22 &     $-$0.05 \\
  ENSO      &     0.42  &     $-$0.22 &     1     &     $-$0.16 \\
  Linear    &     $-$0.24 &     $-$0.05 &     $-$0.16 &     1     \\
  \bottomhline
  \end{tabular}
  \end{table}

Table 3 reports the cross-correlation coefficient matrix among these
four constructors. The cross-correlation coefficients are significantly
smaller than those found in Table~1. In particular, the cross-correlation
coefficients involving the 11\,yr solar cycle constructor with the
other three constructors are very small: $r=0.01$, $r=-0.22$ and
$r=-0.05$, respectively. Thus, this simpler regression model is expected
to be mathematically more robust than that adopted in BS09.

The 1980--2003 period is used to keep the number of fitting parameters
to a minimum. The model (Eq.~\ref{eq:4}) is used to fit three MSU temperature
$T(t)$ records \citep{Christy}: Temperature Lower Troposphere (TLT,
MSU 2); Temperature Middle Troposphere (TMT, MSU 2); Temperature Lower
Stratosphere (TLS, MSU 4). The evaluated regression coefficients are
recorded in Table~4.

Figure~6a shows the three original volcano, solar and ENSO sequences.
Figure~6b shows the three regression models against the MSU temperature
records. Figure~6c shows the reconstructed 11\,yr cycle solar signatures
in the TLT, TMT and TLS records. Finally, Fig.~6d shows the GISS
ModelE reconstruction of the solar signatures from the ground surface
to the lower stratosphere.

The comparison between Fig.~6c and d stresses the striking discrepancy
between the empirical findings and the GISS ModelE predictions for the 11\,yr
solar cycle signatures on climatic records. The empirical analysis shows that
the peak-to-trough amplitude of the response to the 11\,yr solar cycle
globally is estimated by the regression model to be approximately
$0.12$\,$^\circ$C near the earth's surface and rises to 0.3--0.4\,$^\circ$C
at the lower stratosphere. This result agrees with what was found by other
authors \citep{Coughlin,Crooks,Gleisner,Haigh,Labitzke,van
Loon,Scafetta2005,Scafetta2009b,White}. On the contrary, the GISS ModelE
predicts a peak-to-trough amplitude of the climatic response to the solar
cycle globally of $\sim0.03$\,$^\circ$C near the ground, rising to
$\sim0.05$\,$^\circ$C at the lower stratosphere (MSU4). Consequently, the
GISS ModelE climate simulations significantly underestimate the empirical
findings by a factor of $\sim3$ or $4$ for the surface measurements, up to a
factor of $\sim8$ for the lower stratosphere measurements.

%t4
 \begin{table}
  \caption{Values of the regression coefficients used in Eq.~(\ref{eq:4}).
  Units depend on  the original sequences.}
  \center
 \begin{tabular}{lrrr}
 \tophline
  & TLT & TMT & TLS \\
  \hhline
  $\alpha_\mathrm{V}$ & $-$3.18 & $-$2.31 & 8.94 \\
  $\alpha_\mathrm{S}$ & $1.07 \times 10^{-4}$ & $1.25 \times 10^{-4}$ & $2.86 \times 10^{-4}$ \\
  $\alpha_\mathrm{E}$ & 0.131 & 0.139 & 0.0098 \\
   $a$ &0.016 & 0.011& $-$0.027  \\
    $b$ &$-$0.28 & $-$0.28&  $-$0.37 \\
    \bottomhline
    \end{tabular}
   \end{table}

A low response of the climate system to solar changes is not peculiar to the
GISS ModelE alone, but appears to be a common characteristic of present-day
climate models. For example, the predicted peak-to-trough amplitude of the
global surface climate response to the 11\,yr solar cycle is about
$0.025$\,$^\circ$C in Crowley's~(\citeyear{Crowley}) linear upwelling/diffusion energy
balance model; it is about $0.03$\,$^\circ$C in Wigley's MAGICC energy
balance model \citep{Foukal2004,Foukal2006}; it is just a few hundredths of a
degree in several other energy balance models analyzed by \citet{North2004}.

Eventually, in order to correct this situation, other feedback mechanisms and
solar inputs than the total solar irradiance forcing alone should be
incorporated into the climate models as those adopted by the
\citet{IPCC2007}. Possible candidates are a cosmic ray's modulation of the
cloud system that alters the albedo
\citep{Kirkby,Svensmark2007,Svensmark2009}, mechanisms related to UV effects
on the stratosphere and others. For example, \citet{Solomon} estimated that
stratospheric water vapor has largely contributed both to the warming
observed from 1980--2000 (by 30\,\%) and to the slight cooling observed after
2000 (by 25\,\%). This study reinforced the idea that climate change is more
complex than just a reaction to added CO$_{2}$ and a few other anthropogenic
forcings. The causes of stratospheric water vapor variation are not
understood yet. Perhaps stratospheric water vapor is driven by UV solar
irradiance variations through ozone modulation, and works as a climate
feedback to solar variation \citep{Stuber}. Ozone variation may also be
driven by cosmic ray \citep{Lu2009a,Lu2009b}.

%f6
\begin{figure*}[t]
\center
\includegraphics[width=17cm]{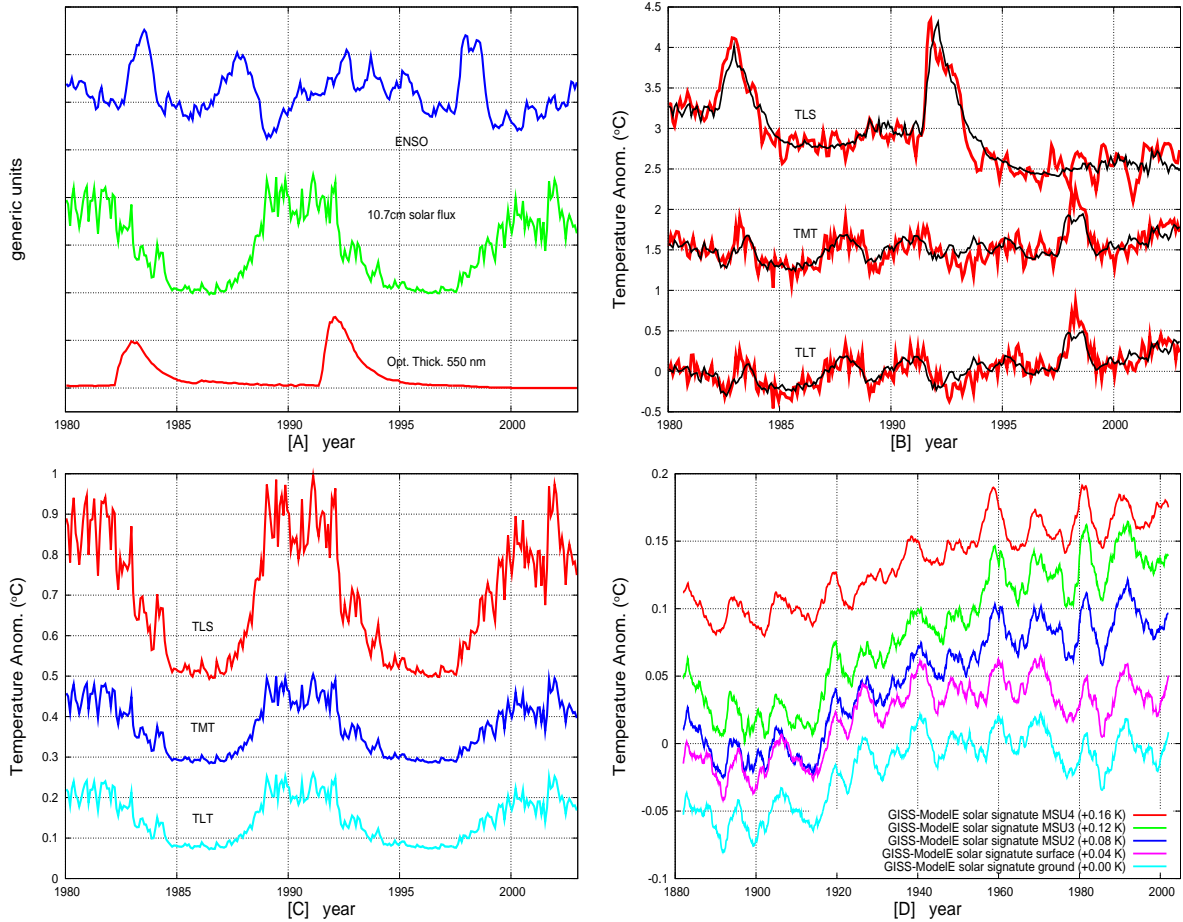}
\caption{\textbf{(A)}~Monthly mean optical thickness at 550\,nm associated to
the volcano signal, the 10.7\,cm solar flux values, and the ENSO MEI signals
used in the regression model of Eq.~(\ref{eq:4}); \textbf{(B)}~Regression
model against MSU temperature records: Temperature Lower Troposphere (TLT,
MSU 2); Temperature Middle Troposphere (TMT, MSU 2); Temperature Lower
Stratosphere (TLS, MSU 4). \textbf{(C)}~Solar Signatures as predicted by the
regression model Eq.~(\ref{eq:4}). \textbf{(D)}~GISS ModelE solar signature
prediction from the ground (bottom) to the lower stratosphere (MSU4) (top).
Regression coefficients are reported in Table~4.}
\end{figure*}

However, BS09 regression model is also not meaningful for another important
physical property. Secular-long climatic sequences cannot be modeled using a
linear regression model that directly adopts as linear predictors the
radiative forcing functions, as done in BS09, because the climate processes
the forcing functions non-linearly by deforming their geometrical shape
through its heat capacity. See the discussion in \citet{Crowley},
\citet{Scafetta2005,Scafetta2006a,Scafetta2006b,Scafetta2007} and
\citet{Scafetta2009b}. Essentially, an input radiative forcing function and
the correspondent modeled temperature output function do not have the same
geometrical shape because each frequency band is processed in different ways
(e.g. high frequencies are damped while low frequencies are stretched), and
multilinear regression models are extremely sensitive to the shape of the
constructors. The same critique applies to \citet{Lean2009}, who also adopted
forcing functions as temperature linear predictors to interpret the 20th
century warming.

The above problem may be circumvented by using a regression model
that uses as predictors theoretical climatic fingerprints of the single
forcing functions once processed by an energy balance model, which
approximately simulate the climate system, as proposed for example
in \citet{Hegerl}. A simple first approximation choice may be a regression
model of the temperature of the type:
\begin{equation}
T(t)=\alpha_{T{_\mathrm{v}}}T_\mathrm{V}(t)+\alpha_{T{_\mathrm{s}}}T_\mathrm{S}(t)+\alpha_{T{_\mathrm{a}}}T_\mathrm{A}(t)+c,\label{eq:model}
\end{equation}
where $T_\mathrm{v}(t)$, $T_\mathrm{S}(t)$ and $T_\mathrm{A}(t)$ are the
outputs of an energy balance model forced with the volcano, solar and
anthropogenic (GHG plus Aerosol) forcing functions, respectively, and $c$ is
a constant. The rationale is the following. Energy balance models provide
just a rough modeling of the real climatic feedbacks and processes involved
in a specific forcing. What the regression model does is estimate signal
amplitudes $\alpha$ (unitless) as scaling factors by which energy balance
model simulations need to be scaled for best agreement with observations
\citep{Hegerl}. This scaling process also makes, in first approximation, the
final results approximately independent of the specific energy balance model
used to produce the constructors. In particular, the scaling factor is
 important for determining a first approximation climatic
contribution of the overall solar variations which, as explained above,
likely present additional forcing (cosmic ray, UV, etc.) functions that
present geometrical similarities to the total solar irradiance forcing
function alone, but that are not explicitly included in the climate models
yet. Indeed, the multiple solar forcings are very likely
quasi-multicollinear, which allows the regression model,
Eq.~(\ref{eq:model}), to approximately estimate, through the scaling factor
$\alpha_{T{_\mathrm{s}}}$, their overall effect by using only a theoretical
climatic fingerprint of one of them.

%f7
\begin{figure*}[t]
\center
\includegraphics[width=17cm]{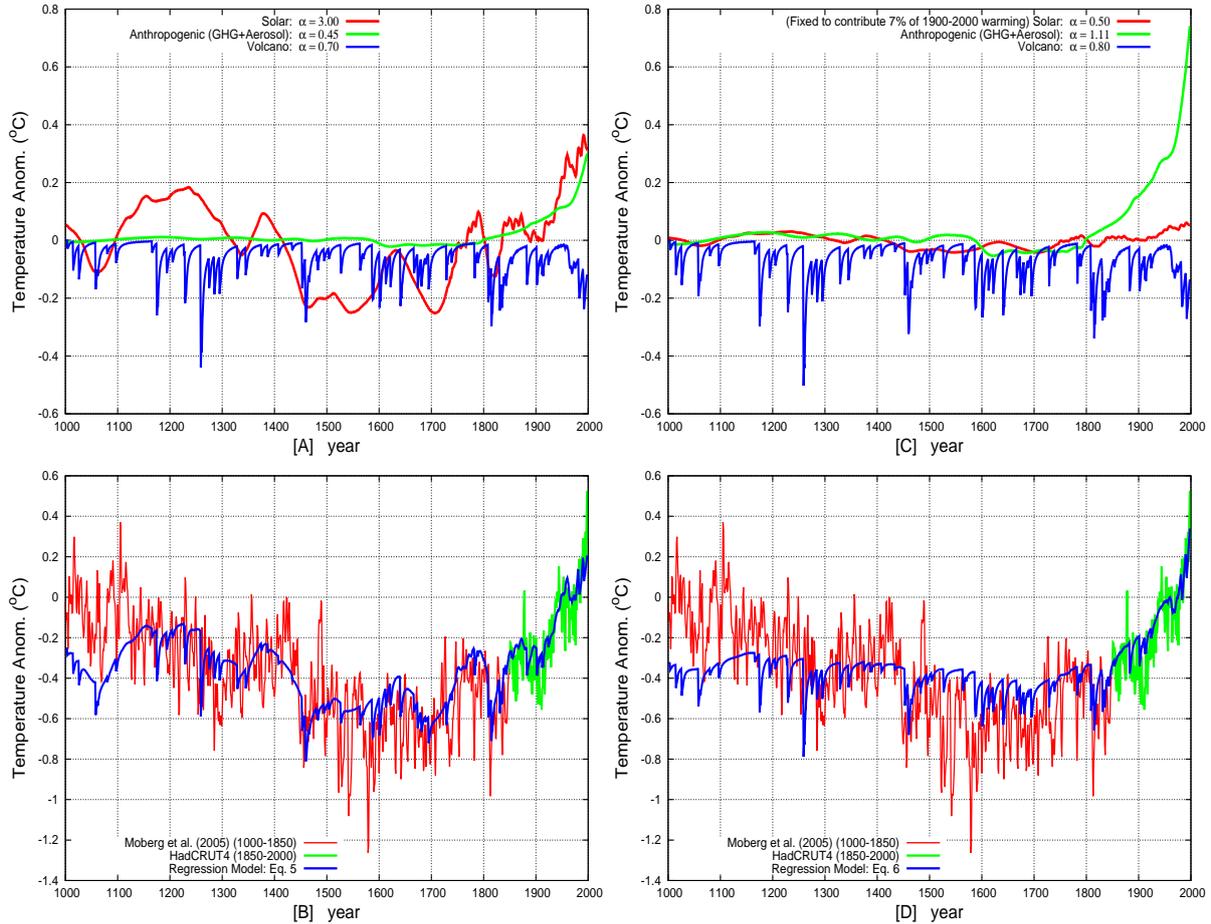}
\caption{\textbf{(A)}~Temperature components obtained with the regression
model, Eq.~(\ref{eq:model}), using as constructors the energy balance model
output functions proposed by \citet{Crowley}. \textbf{(B)}~Comparison between
the model Eq.~(\ref{eq:model}) and an estimate of the global surface
temperature from 1000 to 2000\,AD. The temperature is made of the proxy model
by \citet{Moberg} (1000--1850, red) and by the HadCRUT4 global surface
temperature record (1850--2000, green) \citet{Morice}. \textbf{(C)} and
\textbf{(D)} are like \textbf{(A)} and \textbf{(B)}, but here the restricted
regression model, Eq.~(\ref{eq:model-1}), is used, where the solar
contribution to the 20th century warming is forced to be 7\,\%
($0.06$\,$^\circ$C) of the total ($0.8$\,$^\circ$C), as claimed by \citet{BS09}. Note
that the model, Eq.~(\ref{eq:model-1}), fails to agree with the temperature
before 1750 by not reproducing the great millennial temperature cycle. The regression models appear too rough to properly reconstruct the period 1990-2000.}
\end{figure*}

Because both solar and anthropogenic forcing functions have been increasing
since about 1700 (since the Maunder solar minimum and a cold period of the
Little Ice Age), to reduce the multicollinearity among the constructors, the
regression model should be run against 1000\,yr-long temperature records to
better take advantage of the geometrical orthogonality between the millennial
solar cycle \citep[e.g.: ][]{Bond,Kerr,Ogurtsov,Scafetta2012c} and the GHG
records. Note that the GHG forcing functions show only a small preindustrial
variability and reproduce the shape of a hockey stick \citep{Crowley}. I
observe that this crucial point was also not recognized by \citet[figure
5]{Rohde}, who used a regression model of the temperature from 1750 to 2010,
and also used predictors equivalent to the forcing functions as in BS09.

In the present example I used the output functions produced by the linear
upwelling/diffusion energy balance model from \citet[figure 3A]{Crowley} in
the following way: the volcano output is used as a candidate for the volcano
related constructor; the GHG and Aerosol outputs are summed to obtain a
comprehensive anthropogenic constructor function; the three solar outputs,
which for the 20th century use \citet{Lean2000} solar model, are averaged to
obtain an average solar constructor function.

Note that \citet{Crowley} and \citet{Hegerl} compared their models against
\textit{hockey-stick} temperature reconstructions such as that proposed by
\citet{Mann1999}, which showed a very little preindustrial variability
compared with the post-1900 global warming, and found a relatively small
solar signature on climate. However, since 2005, novel paleoclimatic
temperature reconstructions have demonstrated a far greater preindustrial
variability made of a large millennial cycle with an average cooling from the
Medieval Warm Period to the Little Ice Age of about 0.7\,$^\circ$C
\citep{Moberg,Mann2008,Ljungqvist,Christiansen}; the latter cooling is about
3--4 times greater than what showed by the \textit{hockey-stick} temperature
graphs. The example uses the reconstruction of \citet{Moberg} assumed to
represent a global estimate of the surface temperature, merged in 1850--1900
with the instrumental global surface temperature record HadCRUT4
\citep{Morice}.

The result of the analysis are shown in Fig.~7a and~b. The evaluated scaling
coefficients using Eq.~(\ref{eq:model}) are $\alpha_{T{_\mathrm{v}}}=0.7$;
$\alpha_{T{_\mathrm{s}}}=3.0$; $\alpha_{T{_\mathrm{a}}}=0.45$;
$c=-0.30$\,$^\circ$C. Figure~7a shows the rescaled energy balance model
simulations relative to the three components (volcano, solar and
anthropogenic). Figure~7b shows the model, Eq.~(\ref{eq:model}), against the
chosen temperature record and a good fit is found. According to the proposed
model, from 1900 to 2000 the solar component contributed
$\sim0.35$\,$^\circ$C (44\,\%) of the total $\sim0.8$\,$^\circ$C. However,
this is likely a low estimate because a fraction (perhaps 10--20\,\%) of the
post-1900 GHG increase may have been a climatic feedback to the solar-induced
warming itself through CO$_{2}$ and CH$_{4}$ released by up-welled water
degassing, permafrost melting and other mechanisms. Thus, more likely, the
sun may have contributed at least about 50\,\% of the 20th century warming as
found in other empirical studies
\citep[e.g.:][]{Eichler,Scafetta2007,Scafetta2009b} that more properly
interpreted the climate system response to solar changes using long records
since 1600\,AD, and by taking into account also the scale-by-scale response
of the climate system to solar inputs. Indeed, although the regression model appears too rough, Fig. 7a suggests that solar activity and anthropogenic forcings could have contributed almost equally to the global warming observed from 1900 to 2000.

On the contrary, by comparison, Fig.~7c and~d show what would
have been the situation if the solar contribution to the 20th century
warming were only 7\,\% of the total, that is about $0.06$\,$^\circ$C
against $0.8$\,$^\circ$C, as claimed by BS09 and also by the GISS ModelE
(see Fig.~6d). Here, I forced the solar component of the regression
model to reproduce such a claim, which necessitates a rescaling of
Crowley's solar output $T_\mathrm{S}(t)$ by a factor $\alpha_{T{_\mathrm{s}}}\approx0.5$.
Then, a modification of the regression model of Eq.~(\ref{eq:model})
can be used:
\begin{equation}
T(t)-0.5 \cdot T_\mathrm{S}(t)=\alpha_{T{_\mathrm{v}}}T_\mathrm{V}(t)+\alpha_{T{_\mathrm{a}}}T_\mathrm{A}(t)+c.\label{eq:model-1}
\end{equation}
The three regression coefficients are $\alpha_{T{_\mathrm{v}}}=0.8$,
$\alpha_{T{_\mathrm{a}}}=1.1$ and $c=-0.32$\,$^\circ$C. As Fig.~7c shows, in
this case almost the entire 20th century global warming would be interpreted
as due to anthropogenic forcings, as all general circulation models of the
\citet{IPCC2007} and also \citet{Lean2009} have claimed. However, as Fig.~7d
clearly highlights, the same model, Eq.~(\ref{eq:model-1}), fails to
reproduce the data before 1750 by missing the great millennial oscillation,
generating both the Medieval Warm Period (1000--1400) and the Little Ice Age
(1400--1750). Equation~(\ref{eq:model-1}) just reproduces a hockey-stick
shape that would only agree well with the outdated paleoclimatic
reconstruction by \citet{Mann1999}, as also originally found by
\citet{Crowley}.

It is possible to observe that a good agreement between the model,
Eq.~(\ref{eq:model-1}), and the data since 1750 exists, which is the same
result found by \citet[figure 5]{Rohde} with another regression model. These
authors concluded that almost all warming since 1750 was induced by
anthropogenic forcing. However, \citet{Rohde} result is also not robust
because their used 260\,yr interval (1750--2010) is too short a period,
during which both the solar and the anthropogenic forcing functions are
collinear (both increased); a regression model therefore cannot properly
separate the two signals.

In conclusion, it is evident that the large preindustrial millennial
variability shown by recent paleoclimatic temperature reconstructions implies
that the sun has a strong effect on the climate system, and its real
contribution to the 20th century warming is likely about 50\,\% of the total
observed warming. This estimate is clearly incompatible with BS09's estimate
of a solar contribution limited to a mere 7\,\% of the 20th century global
warming. The result is also indirectly confirmed by the results depicted in
Fig.~6, demonstrating that GISS ModelE severely underestimates the solar
fingerprints on climatic records by a large factor. For equivalent reasons,
by claiming a very small solar effect on climate, the general circulation
models used by the \citet{IPCC2007} and the regression models proposed by
\citet{Rohde} and \citet{Lean2009}, would  be physically compatible only
with the outdated hockey-stick paleoclimatic temperature graphs \citep[e.g.
][]{Crowley,Mann1999} if they were extend back to 1000\,AD. However, by doing
so, they would fail to reproduce the far larger (by a factor of 3 to 4, at
least) preindustrial climatic variability revealed by the most recent
paleoclimatic temperature reconstruction \citep[e.g.:
][]{Christiansen,Kobashi2013,Ljungqvist,Mann2008,Moberg}.

\section{Maximum overlap discrete wavelet transform,  Gibbs artifacts and boundary methods}

A technique commonly used to extract structure from complex time series
is the maximum overlap discrete wavelet transform (MODWT) multiresolution
analysis (MRA) \citep{Percival}. This methodology decomposes a signal
$X(t)$ at the $J$-{th} order as follows:
\begin{equation}
X(t)=S_{J}(t)+\sum_{j=1}^{J}D_{j}(t),\label{eq:5}
\end{equation}
where $S_{J}(t)$ works as a low-pass filter and captures the smooth
modulation of the data with time scales larger than $2^{J+1}$ units
of the time interval $\Delta t$ at which the data are sampled. The
detail function $D_{j}(t)$ works as a band-pass filter and captures
local variation with periods approximately ranging from $2^{j}\Delta t$
to $2^{j+1}\Delta t$. The technique can be used to model the climatic
response to different temporal scales of the solar forcing used later
to combine the results to obtain a temperature signature induced by
solar forcing as proposed in \citet{Scafetta2005,Scafetta2006a,Scafetta2006b}.

%f8
\begin{figure*}[t]
\center
\includegraphics[width=14cm]{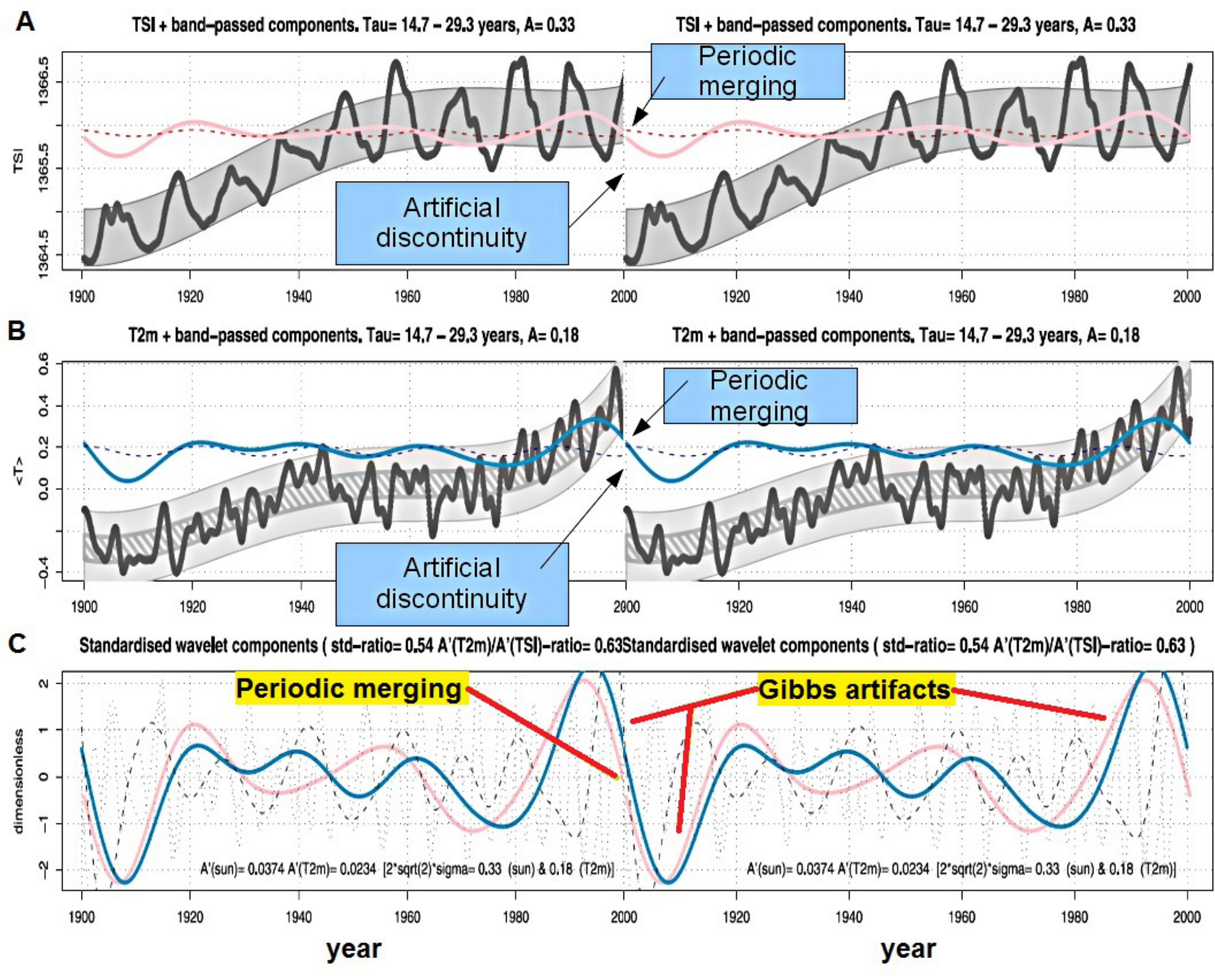}
\caption{Reproduction and comments of \citet{BS09}'s figure 4 applying the MODWT to:
\textbf{(A)}~a total solar irradiance model \citep{Lean2000};
\textbf{(B)}~the GISTEMP global surface temperature \citep{Hansen2001}. The bottom
panel~\textbf{(C)} depicts the MODWT decomposed curves (blue and pink)
at a time scale of about $\tau=22$ yr. The figure plots the original
figure twice, side by side to demonstrate that MODWT was applied improperly
with the periodic boundary method.}
\end{figure*}

%f9
\begin{figure}[t]
\center
\includegraphics[width=8.5cm]{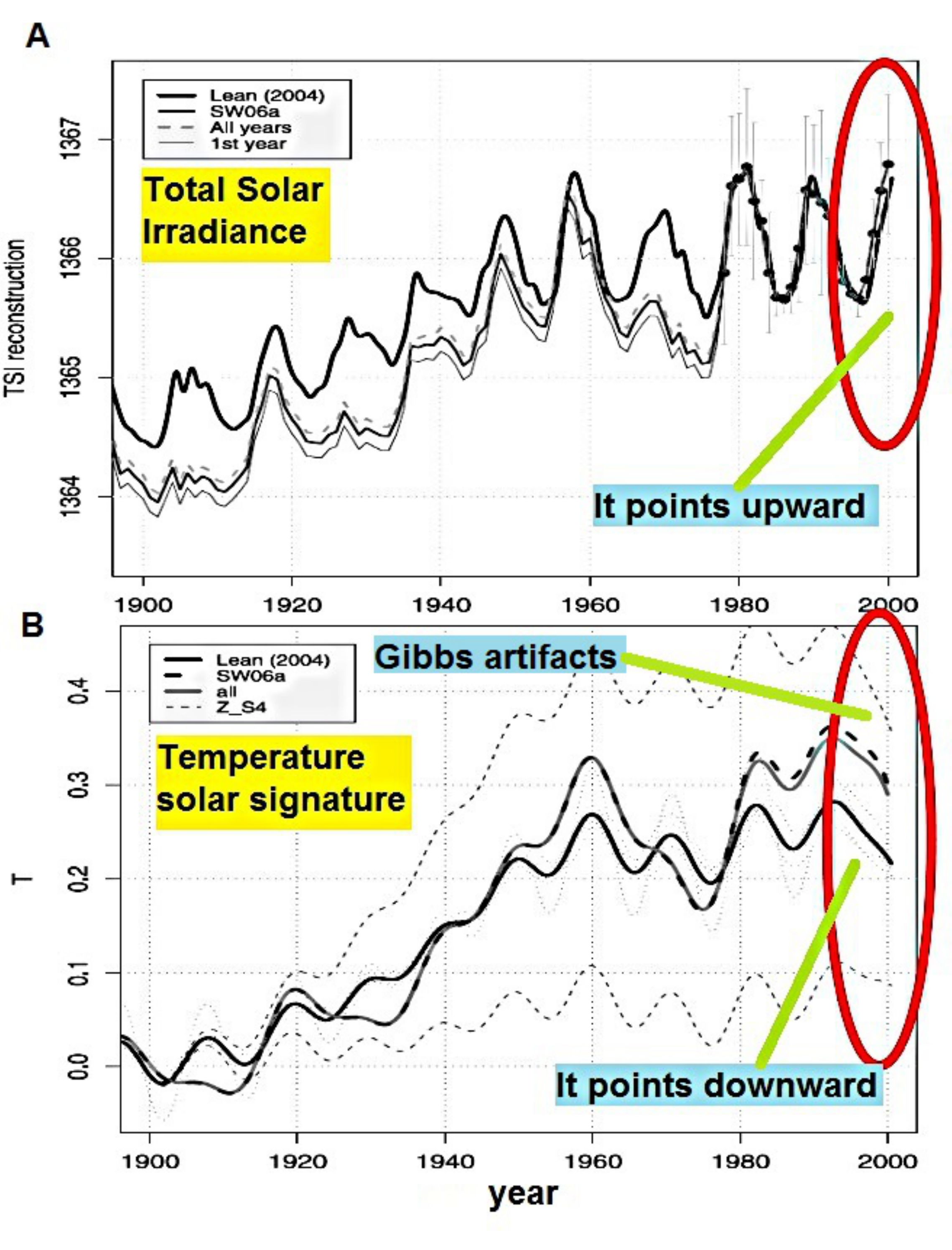}
\caption{Reproduction and comment of \citet{BS09}'s figure 6~\textbf{(A)} and figure~7~\textbf{(B)}.
The upper panel shows a total solar irradiance model \citep{Lean2000};
the bottom panel shows its temperature signature produced with the
MODWT decomposition at scales larger than the decadal one. The red
circles highlight the physical incongruity of the misapplication of
the MODWT method by showing that despite the increasing solar activity
(upper panel) the wavelet processed curve points downward because
of Gibbs artifacts.}
\end{figure}

However, the MODWT technique needs to be applied with care because
the MODWT pyramidal algorithm is periodic \citep{Percival}. This
characteristic implies that to properly decompose a non-stationary
time series, such as solar and temperature records, there is the need
of doubling the original series by reflecting it in such a way that
the two extremes of the new double sequence are periodically continuous:
that is, if the original sequence runs from A to B (let us indicate
it as ``A-B''), it must be doubled to form a sequence of the type
``A-BB-A'', which is periodic at the extremes. This boundary method
is called ``reflection''. If this trick is not applied and the original
sequence is processed with the default periodic boundary method, MODWT
interprets the sequence as ``A-BA-B'', and models the ``BA'' discontinuities
at the extremes by producing Gibbs ringing artifacts that invalidate
the analysis and its physical interpretation.

A serious misapplication of the MODWT methodology is also found in
\citet{BS09}, where they questioned the MODWT results of temperature and solar
records found in \citet{Scafetta2005,Scafetta2006a,Scafetta2006b}
because they were not able to reproduce them. However, as I will demonstrate
below, BS09 misapplied the MODWT by using the periodic boundary method
instead of the required reflection one. This error could have been
easily recognized by a careful analysis of their results, which were
weird. Let us discuss the case.

Figure~8 reproduces BS09's figure 4, which decomposes with MODWT both a total
solar irradiance record \citep{Lean2000} and the GISTEMP global surface
temperature record \citep{Hansen2001}. Figure~8, however, plots BS09's figure 4
twice, by merging the 2000 border with the 1900 border side-by-side. As
evident in the figure, the original sequences present discontinuities at the
borders of the type ``A-BA-B'' due to their upward trend. However, the MODWT
decomposed curves, that is, the pink and blue curves are continuous at the
borders. This pattern is generated by the MODWT when the default periodic
boundary method is applied. Consequently, very serious Gibbs artifacts are
observed in the decomposed curves, as evident in the large oscillations
present in the pink and blue curves in proximity of the borders in 1900 and
in 2000. These artifacts are also evident in the bottom panel of Fig.~8.

The consequences of the error are quite serious. The Gibbs artifacts induce a
large artificial volatility in the decomposed components signals. Moreover,
they generated a serious physical incongruity highlighted in Fig.~9a and~b.
These figures reproduce BS09's figures 6 and 7, respectively. Here, a total
solar irradiance model \citep{Lean2000} (Fig.~9a) and its temperature
signature reconstruction using the MODWT decomposition methodology of
\citet{Scafetta2006a} (Fig.~9b) are depicted, respectively. Figure~9a shows
that with the MODWT methodology, the solar contribution to the 20th century
global warming is about $0.3$\,$^\circ$C (38\,\%) of the total warming.
However, the exact result would be larger if the MODWT were not misapplied
and would have fully confirmed \citet{Scafetta2006a,Scafetta2006b}. In fact, the added
red circles in Fig.~9 highlight that the solar activity increased from 1995
to 2000 (Fig.~8a), while its temperature reconstructed signature (Fig.~9b)
points downward during the same period, which is unphysical. This pattern was
due to the fact that the algorithm, as applied by BS09, processed the signal
with the default periodic boundary method and generated large Gibbs artifacts
trying to merge the starting and the ending points of the record and,
consequently, bent the last decade of the reconstructed solar climatic
signature downward. Note that in \citet[compare figures 2 and
3]{Scafetta2006a}, where the MODWT was applied correctly using the reflection
boundary method, this physical incongruity does not exist. \citet{Scafetta2006a,Scafetta2006b,Scafetta2007} and \citet{Scafetta2009b} are consistent also with the results discussed in Sect. 3 and Figure 7 where  it was determined that the sun contributed about 50\% of the global warming from 1900 to 2000 using an independent methodology.

Let us discuss how to correctly apply the MODWT methodology as used in
\citet{Scafetta2005,Scafetta2006a} to capture the 11\,yr and 22\,yr solar
cycle signatures. In addition to the reflection method for mathematical
purposes, MODWT methodology has to be used under the following two conditions
for physical reasons: (1)~the data record needs to be resampled in such a way
that the center of the wavelet band-pass filter is located exactly on the 11
and 22\,yr solar cycles, which are the frequencies of interest; (2)~a
reasonable choice of the year when the reflection is made, that is, the year
2002--2003 when the sun experienced a maximum for both the 11\,yr and 22\,yr
cycles to further reduce a problem of discontinuity in the derivative at the
border because there is the need to apply MODWT with the reflection method.

%f10
\begin{figure*}[t]
\center
\includegraphics[width=13cm]{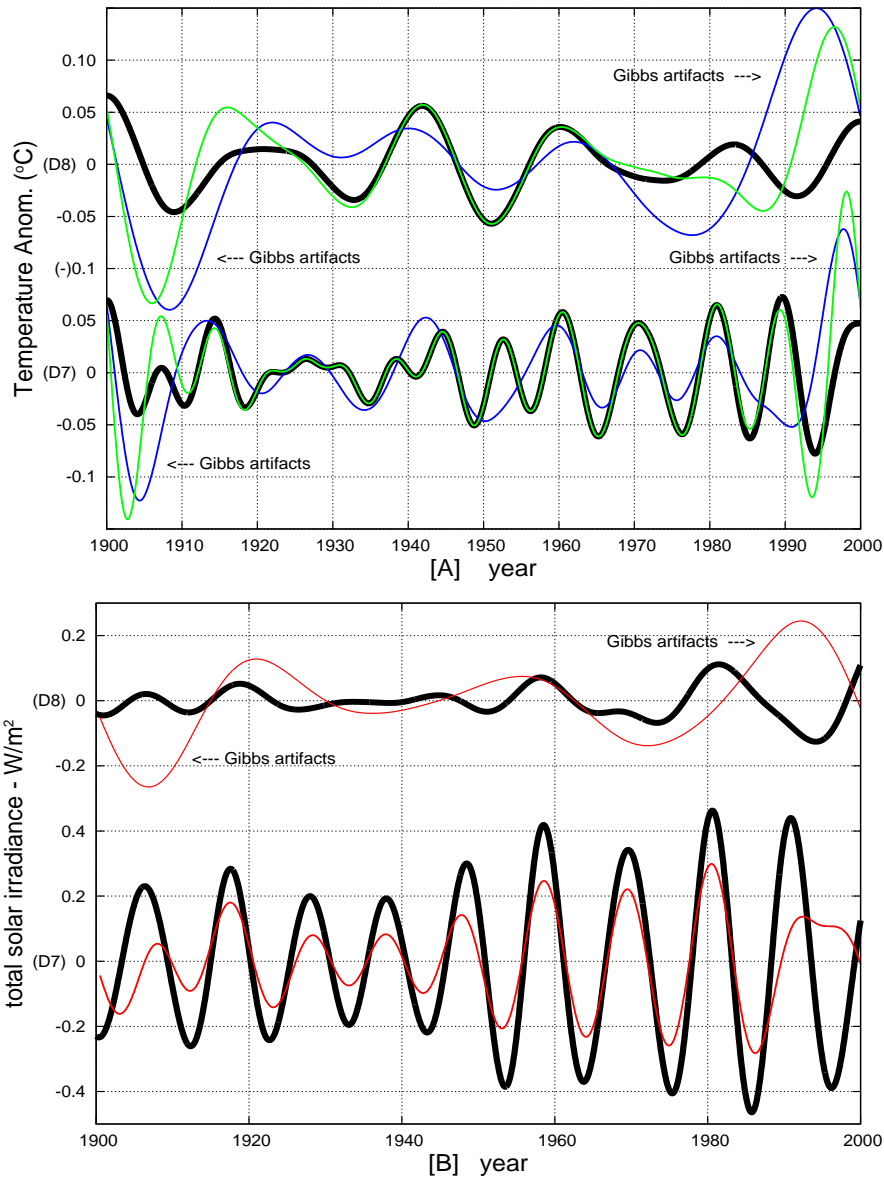}
\caption{\textbf{(A)}~MODWT of the GISTEMP dataset \citep{Hansen2001} from 1900
to 1999. (Top) Detail curve $D_{8}$; (Bottom) Detail curve $D_{7}$.
The thick black curves are the correct $D_{7}$ and $D_{8}$ wavelet
detail curves obtained with reflection boundary condition and the
correct time step of $\Delta t=0.6875$\,month. The thin blue lines
``exactly'' correspond to the curves depicted in Fig.~8 in blue
(central panel) and blue and dash (bottom panel). The thin green lines
are obtained using the periodic method and the correct time step of
$\Delta t=0.6875$\,month. \textbf{(B)}~MODWT of the \citep{Lean2000}
TSI from 1900 to 1999. (Top) Detail curve $D_{8}$; (Bottom) Detail
curve $D_{7}$. The thick black curves are the correct $D_{7}$ and
$D_{8}$ wavelet detail curves obtained with reflection boundary condition
and the correct time step of $\Delta t=0.6875$\,month. The thin
red curve in the upper panel ``exactly'' corresponds to the curves
reported in Fig.~8 in pink (upper and bottom panels) that were obtained
by using the cyclical periodic boundary condition and the incorrect
time step of $\Delta t=1$\,month.}
\end{figure*}

Point (1) was accomplished by observing that the 11\,yr cycle (132
months) would fall within the frequency band captured by the wavelet
detail $D_{7}(t)$ corresponding to the band between $2^{7}=128$
and $2^{8}=256$ months, that is from 10.7 to 21.3\,yr. Thus,
by adopting the monthly sampling the 11\,yr cycle would not be centered
in $D_{7}(t)$ and the 22\,yr would not be centered in the wavelet
detail $D_{8}(t)$. This would cause an excessive splitting of the
11\,yr modulation between the adjacent details curve $D_{6}(t)$
and $D_{7}(t)$, and of the 22\,yr modulation between the adjacent
detail curves $D_{7}(t)$ and $D_{8}(t)$. Consequently, to optimize
the filter it was necessary to adjust the time step of the time sequence
in such a way that the wavelet detail curves fell exactly in the middle
of the 11\,yr and 22\,yr cycles. This was done by adjusting the
time sampling of the record to $\Delta t=132/192=0.6875$ month or
to $\Delta t=11/12=0.9167$\,yr whether the original sequence has
a monthly or annual resolution, respectively. This time step adjustment
was accomplished with a simple linear interpolation of the original
sequence. With the new resolution $\Delta t=0.6875$ month, the detail
curve $D_{7}(t)$ would cover the timescales 7.3--14.7\,yr (median
11\,yr), and the detail curve $D_{8}(t)$ would cover the timescales
14.7--29.3\,yr (median 22\,yr). This time step is necessary to
optimally extract the 11\,yr and 22\,yr modulations from the data.

%t5
\begin{table*}
\caption{Comparison of the amplitudes of the detail curves shown in Fig.~10
calculated by fitting with function $f(x,P)=0.5A\cos(2\pi(x-T)/P)$,
whe the period $P$ is chosen to be 11\,yr for $D_{7}$ or for 22\,yr
for $D_{8}$. First column shows the amplitudes $A$ for the
period 1980--2000 using the correct reflection method and the optimized
time sampling $\Delta t=0.6875$\,month. Second column shows the
amplitudes $A$ for the period 1980--2000 using the incorrect periodic
method and the non-optimized time sampling $\Delta t=1$\,month.
Third column shows the values as calculated in \citet{BS09}, which are more
similar to those listed in column 2.}
\center
\begin{tabular}{llll}
\tophline
 & Reflection (correct) & Periodic (incorrect) & BS09\\
 \hhline
$A_\mathrm{8,temp}$ & $0.057\pm0.0015$\,$^\circ$C & $0.193\pm0.002$\,$^\circ$C & $0.18$\,$^\circ$C \\
$A_\mathrm{8,sun}$  & $0.226\pm0.003$\,W\,m$^{-2}$ & $0.266\pm0.0025$\,W\,m$^{-2}$ & $0.32$\,W\,m$^{-2}$ \\
$A_\mathrm{8,temp}/A_\mathrm{8,sun}$  & $0.252\pm0.010$\,$^\circ$C\,W$^{-1}$\,m$^2$ & $0.726\pm0.014$\,$^\circ$C\,W$^{-1}$\,m$^2$ & $0.56$\,$^\circ$C\,W$^{-1}$\,m$^2$ \\
$A_\mathrm{7,temp}$  & $0.112\pm0.005$\,$^\circ$C & $0.077\pm0.010$\,$^\circ$C & $0.14$\,$^\circ$C \\
$A_\mathrm{7,sun}$ & $0.872\pm0.008$\,W\,m$^{-2}$ & $0.292\pm0.021$\,W\,m$^{-2}$ & $0.45$\,W\,m$^{-2}$ \\
$A_\mathrm{7,temp}/A_\mathrm{7,sun}$  & $0.128\pm0.007$\,$^\circ$C\,W$^{-1}$\,m$^2$ & $0.264\pm0.053$\,$^\circ$C\,W$^{-1}$\,m$^2$ & $0.311$\,$^\circ$C\,W$^{-1}$\,m$^2$ \\
\bottomhline
\end{tabular}
\end{table*}

Figure~10a shows the MODWT decomposition of the GISTEMP temperature
record from 1900 to 2000. The thick black curves are the $D_{7}$
(bottom) and $D_{8}$ (top) wavelet detail curves obtained with the
reflection method and the correct time step $\Delta t=0.6875$\,month.
A second set of curves are also depicted in Fig.~10a and these are
obtained from the same data, but using cyclic boundary conditions
and with $\Delta t=1$\,month (blue) and $\Delta t=0.6875$\,month
(green). The blue and green curves of Fig.~10a show substantial
Gibbs ringing artifacts that are so serious that they even cause an
inversion of the bending of the curve. The blue curves exactly correspond
to the blue (middle panel) and to the blue and dash (bottom panel)
curves depicted in Fig.~8, as calculated by BS09 where these Gibbs
artifacts were mistakenly identified as anomalous temperature and
solar signatures. The re-analysis also demonstrates that the calculations
by BS09 used the $\Delta t=1$\,month resolution, contrary to what
they report in their figure 4.

The consequences of the error of using MODWT with the default periodic
method are significant. For example, the MODWT detail curves were
used to estimate the average peak-to-trough amplitudes $A$ of the
oscillations from 1980 to 2000. The blue curves for $D_{8}$ gives
$A_\mathrm{8,temp}\approx0.19$\,$^\circ$C, while the value determined using
the correct analysis is $A_\mathrm{8,temp}\approx0.06$\,$^\circ$C corresponding
to the one obtained in \citet{Scafetta2005}. Note also that from
1980 to 2000 the blue curve referring to $D_{8}$ is concave while
the correct curve (black) is convex. Analogous problem referring to
$D_{8}$ is shown in Fig.~10b that analyzes the total solar irradiance
of \citet{Lean2000}. Again the thick black curves are the $D_{7}$
(bottom) and $D_{8}$ (top) wavelet detail curves obtained with the
reflection method and the centered time step $\Delta t=0.6875$\,month.
The thin red lines correspond to the pink curves of Fig.~8. These
latter curves are quite different from the black curves because of
Gibbs artifacts and because the $\Delta t=1$ month resolution was
used. In particular, notice the visibly smaller amplitude of the 11\,yr
solar cycles relative to the correct black curve: with BS09's methodology
one would find $A_\mathrm{7,sun}=0.3$\,W\,m$^{-2}$ while the correct amplitude
is significantly larger $A_\mathrm{7,sun}\approx0.9$\,W\,m$^{-2}$ since 1980,
as found in \citet{Scafetta2005}. Moreover, as Fig.~4a clearly
shows, from 1900 to 2000 the amplitude of the 11-year solar cycle
varies from $0.5$ to $1.3$ about W\,m$^{-2}$ and this range is clearly
inconsistent with the average amplitude of $0.45$\,W\,m$^{-2}$ as calculated
in BS09. The various amplitudes  are listed in Table~5, which
also highlights the abnormal results obtained in BS09.

In summary, MODWT requires: (1)~the reflection method; (2)~the sequences
should be sampled at specific optimized time intervals that depend on the
specific application; (3)~for optimal results the borders need to be chosen
to avoid discontinuities in the first derivative at the time scales of
interest. For example, \citet{Scafetta2005} used the period 1980--2002
because the 11\,yr and 22\,yr solar cycles would approximately have been at
their maximum. The latter point is important because the reflection method
gives optimized results when the derivative at the borders approaches zero.
On the contrary, choosing the default periodic method and using sequences
sampled at generic time intervals and generic borders, as done in BS09, was
demonstrated here to yield results contaminated by significant artifacts.

There are other claims raised in BS09 as those based on GISSModelE and GISS CTL simulations
simulations, etc. However, in Sect.~3 it has been also demonstrated that these
simulations do not reproduce the solar and astronomical signatures on the
climate at multiple time scales \citep[see also
][]{Scafetta2010,Scafetta2012b} and would eventually agree only with the outdated hockey-stick temperature reconstructions such as  those proposed by \citet{Mann1999}. Therefore,  BS09's additional arguments are of limited utility (1) because those computer simulations appear to seriously underestimate the solar signature on climate and (2) because, in any case, BS09 misapplied the MODWT methodology to analyze them.

\conclusions

In this paper I have discussed a few typical examples where time series
methodologies used to analyze climatic records have been misapplied.
The chosen examples address relatively simple situations that yielded
severe physical misinterpretations that, perhaps, could have been
easily avoided.

A first example addressed the problem of how to estimate accelerations in
tide gauge records. It has been shown that to properly interpret the tide
gauge record of New York City it is necessary to plot all available data
since 1856, as done in Fig.~3b. In this way the existence of a quasi 60\,yr
oscillation, which is evident in the global sea level record since 1700,
becomes quite manifest. This pattern suggests a very different interpretation
than that proposed for example in \citet{Sallenger} or in \citet{Boon}. A significantly
smaller and less alarming secular acceleration in NYC was found:
$a=0.006\pm0.005$\,mm\,yr$^{-2}$ against \citet{Sallenger}'s 1950--2009 and 1970--2009
accelerations $a=0.044\pm0.03$\,mm\,yr$^{-2}$ and
$a=0.13\pm0.09$\,mm\,yr$^{-2}$, respectively, or Boon's~(\citeyear{Boon}) 1969--2011
acceleration $a=0.20\pm0.07$\,mm\,yr$^{-2}$. These large accelerations simply
refer to the bending of the quasi 60\,yr natural oscillation present in this
record: see \citet{Scafetta2013b} for additional details. Thus, in NYC a more
realistic sea level rise projection  from 2000 to 2100 would be about $350\pm30$\,mm instead of
$1130\pm480$\,mm calculated with \citet{Sallenger}'s method using the 1970--2009
quadratic polynomial fit or $1550\pm400$\,mm calculated with
Boon's~(\citeyear{Boon})
method using the 1969--2011 quadratic polynomial fit. Moreover, as Fig.~1a
shows, by plotting the NYC tide gauge record only since 1950 (compare
Figs.~1a and 3b), \citet{Sallenger} has somehow obscured the real dynamics of this
record; the same critique would be even more valid for \citet[figures
5--8]{Boon} who plotted tide gauge records only starting in 1969.

Global sea level may rise significantly more if during the 21st century the
temperature increases abnormally by several degrees Celsius, as current
general circulation models have projected \citep{Morice}. However, as
demonstrated in Sect.~3 (e.g. Figs.~6 and~7) typical climate models used for
these projections appear to significantly overestimate the anthropogenic
warming effect on climate and underestimate the solar effect. Solar activity
is projected to decrease during the following decades and may add a cooling
component to the climate \citep{Scafetta2012c,Scafetta2013}. As a
consequence, it is very likely that the 21st century global temperature
projections are too high, as also demonstrated in \citet{Scafetta2012b}.
Because the global sea level record presents a 1700--1900 preindustrial
period acceleration compatible with the 1900--2000 industrial period
acceleration ($a=\sim0.01$\,mm\,yr$^{-2}$ in both cases), there is no clear
evidence that anthropogenic forcings have drastically increased the sea level
acceleration during the 20th century. Thus, anthropogenic forcings may not
drastically increase the sea level during the 21th century either. The
1700--2000 global sea level is projected to rise about $277\pm8$\,mm from
2000 to 2100 as shown in Fig.~2a.

A second example addressed more extensively the problem of how to deal with
multilinear regression models. Multilinear regression models are very
powerful, but they need to be used with care to avoid multicollinearity among
the constructors yielding meaningless physical interpretations. It has been
demonstrated that the 10-constructor multi-linear regression model adopted in
\citet{BS09} to interpret the 20th century global warming and to conclude that the
sun contributed only 7\,\% of the 20th century warming is not robust because:
(1) the used predictors are multicollinear and (2) the climate is not a
linear superposition of the forcing functions themselves. About the latter
point, it is evident that if the climate system could be interpreted as a
mere linear superposition of forcing functions (as also done in \citet{Lean2009}), there would be no need to use climate models in the first place.

To demonstrate the serious artifacts generated by regression analyses in
multicollinearity cases, I showed that by eliminating the predictor claimed
to be the most responsible for the observed global warming from 1900 to 2000,
that is the well-mixed greenhouse gas forcing function, the regression model
was still able to reconstruct equally well the temperature record by using
the other 9 constructors. By using the regression model in a more appropriate
way, that is, by restricting the analysis to the 1980--2003 period when the
data are more accurate and using only orthogonal constructors, it was
demonstrated that the GISS ModelE severely underestimates the solar effect on
climate by a 3-to-8 factor, as shown in Fig.~6. By using a more physically
based regression model (Eq.~\ref{eq:model}) it was also demonstrated that the
large preindustrial temperature variability shown in recent paleoclimatic
temperature reconstructions since the Medieval Warm Period implies that the
sun has a strong effect on climate change and likely contributed about 50\,\%
of the 20th century warming, as found in numerous Scafetta's papers
\citep{Scafetta2006a,Scafetta2006b,Scafetta2007,Scafetta2009b,Scafetta2010,Scafetta2012a,Scafetta2012b}
and by numerous other authors \citep[e.g.:
][]{Eichler,Hoyt,Kirkby,Kobashi2013,Soon2005,Soon2011,Soon2013,Svensmark2007}.
These results contradict \citet{BS09}, \citet{IPCC2007}, \citet{Lean2009} and
Rohde et al.'s~(\citeyear{Rohde}) results that the sun contributed little (less than 10\,\%) to
the $0.8$\,$^\circ$C global warming observed from 1900 to 2000. In fact, such
a low solar contribution would only be consistent with the geometrical
patterns present in outdated hockey-stick temperature
reconstructions \citep[e.g. those proposed in: ][]{Crowley,Mann1999}, as
shown in Fig.~7.

A third example addressed the problem of how to deal with scale-by-scale
wavelet decomposition methodologies, which are very useful to interpret
dynamical details in geophysical records. Evidently, there is a need to
properly take into account the mathematical properties of the methodology to
avoid embarrassing artifacts and physical incongruities as those generated by
Gibbs artifacts. For example, \citet{BS09} findings that the solar activity increase
from 1995 to 2000 has induced a cooling on the global climate and their
failure to reproduce the results of
\citet{Scafetta2005,Scafetta2006a,Scafetta2006b} were just artifacts due to
an improper application of the MODWT technique. \citet{BS09} erroneously applied
MODWT with the default period method instead of using the reflection method
as demonstrated in Figs.~8--10. The error in applying correctly the
decomposition methodology also produced abnormally large uncertainties in
their results. I have spent some time to detail how to use this technique for
the benefit of the readers interested in properly applying it.

Highlighting these kinds of problems is important in science. In fact, while
errors in scientific research are sometimes possible and unavoidable, what
most harms the scientific progress is the persistence and propagation of the
errors. This happens when other scientists uncritically cite and use the
flawed results to interpret alternative data, which yields further
misinterpretations. This evidently delays scientific progress and may damage
society as well.

\begin{acknowledgements}
The author would like to thank the Editor and the two referees for useful and constructive comments.
The author thanks Mr. Roger Tattersall for encouragement and suggestions.
\end{acknowledgements}

\end{document}